\documentclass[english,aps,amssymb,pra,showpacs,preprintnumbers,twocolumn]{revtex4}
\pdfoutput=1
\usepackage[T1]{fontenc}
\usepackage[latin9]{inputenc}
\usepackage{amsmath}
\usepackage{graphicx}
\usepackage{amssymb}
\usepackage{esint}
\usepackage{times}

\makeatletter
\@ifundefined{textcolor}{}
{
 \definecolor{WHITE}{gray}{1}
 \definecolor{RED}{rgb}{1,0,0}
 \definecolor{GREEN}{rgb}{0,1,0}
 \definecolor{BLUE}{rgb}{0,0,1}
 \definecolor{CYAN}{cmyk}{1,0,0,0}
 \definecolor{MAGENTA}{cmyk}{0,1,0,0}
 \definecolor{YELLOW}{cmyk}{0,0,1,0}
 }

\renewcommand{\phi}{\varphi}
\renewcommand{\epsilon}{\varepsilon}

\usepackage{epsfig}
\usepackage{babel}

\begin{document}

\title[Ground state cooling of mechanical motion in the unresolved sideband regime by use of optomechanically induced transparency]{Ground state cooling of mechanical motion in the unresolved sideband regime\\ by use of optomechanically induced transparency}

\author{Teemu Ojanen}

\affiliation{Low Temperature Laboratory (OVLL), Aalto University, P.~O.~Box 15100,
FI-00076 AALTO, Finland}

\author{Kjetil B{\o}rkje}
\affiliation{Niels Bohr Institute, University of Copenhagen, Blegdamsvej 17, DK-2100 Copenhagen, Denmark}

\date{\today}
\begin{abstract}
We present a scheme for cooling mechanical motion to the ground state in an optomechanical system. Unlike standard sideband cooling, this scheme applies to the so-called unresolved sideband regime, where the resonance frequency of the mechanical mode is much smaller than the cavity linewidth. Ground state cooling becomes possible when assuming the presence of an additional, auxiliary mechanical mode and exploiting the effect of optomechanically induced transparency. We first consider a system where one optical cavity interacts with two mechanical modes, and show that ground state cooling of the unresolved mechanical mode is possible when the auxiliary mode is in the resolved sideband regime. We then present a modified setup involving two cavity modes, where both mechanical modes are allowed to be in the unresolved sideband regime. 
\end{abstract}

\maketitle
\bigskip{}

\section{Introduction}

Optomechanical systems, where electromagnetic radiation and mechanical motion interact, hold promise for advances in areas such as sensing and information technology \cite{Aspelmeyer2013}. They also offer a novel platform to study quantum behaviour in massive, mesoscopic systems. The canonical optomechanical system consists of an optical Fabry-P\'{e}rot cavity with one movable mirror, such that the cavity resonance frequency is parametrically dependent on the position of the mirror. This type of interaction has been realized in many widely different ways, which have been extensively reviewed in Ref.~\cite{Aspelmeyer2013}.



In order to have quantum control over the mechanical motion, one needs to overcome the thermal fluctuations that inevitably follow from interactions with the environment. Since most nano- and micromechanical resonators typically have resonance frequencies below 10 MHz, placing the system in a dilution refrigerator can be insufficient to achieve ground state cooling of the mechanical motion. To overcome this, laser cooling techniques have been developed. Passive sideband cooling, previously used to cool the motion of trapped atoms \cite{Diedrich1989PRL}, has recently been successfully applied to cool the motion of mesoscopic objects to the quantum ground state \cite{Teufel2011Nature,Chan2011Nature}. The recent observations of ponderomotive squeezing of light \cite{Brooks2012Nature,Purdy2013PRX,Safavi-Naeini2013Nature} and of  optomechanically induced transparency (OMIT) \cite{Agarwal2010PRA,Weis2010Science,Teufel2011Nature_2,Safavi-Naeini2011Nature,Karuza2013PRA} are other important experimental achievements, both of which are relevant to the topic of this article.

Cavity-assisted sideband cooling can reach the mechanical ground state only in the resolved sideband regime where the frequency of the mechanical mode $\omega_1$ and the cavity linewidth $\kappa$ satisfy $\omega_1 > \kappa$ \cite{Wilson-Rae2007PRL,Marquardt2007PRL}. This condition ensures that the rate of Stokes scattering of photons, which creates mechanical vibration quanta (phonons), is suppressed relative to the rate of anti-Stokes scattering, where phonons are destroyed. Preparing the mechanical oscillator in the ground state is then possible, at least in principle, for sufficiently strong optical driving. While many such sideband-resolved systems have been realized, the criterion $\omega_1 > \kappa$ typically requires optical cavities with very high-finesse and it limits the size and mass of the mechanical oscillator to be cooled. It is therefore desirable to develop techniques that are able to circumvent this restriction while still allowing ground state cooling. This is the main motivation for the work presented here.            

In this article, we present a passive cooling scheme relevant to cavity optomechanical systems which allows for ground state cooling of mechanical motion even when $\omega_1 \ll \kappa$. The scheme assumes the presence of an auxiliary mechanical mode with frequency $\omega_2$, as well as an auxiliary optical drive in addition to the drive responsible for cooling. The two mechanical modes can either belong to two separate physical objects, or they can be different normal modes of the same object. For a particular choice of drive frequencies, the interaction of the auxiliary mode with the cavity field can lead to a suppression of Stokes scattering processes that otherwise would hinder ground state cooling of the original mechanical mode. The origin of this suppression can be thought of as optomechanical squeezing of the photon number fluctuations in the cavity. We first show that ground state cooling of the original mode is possible when the auxiliary mechanical mode is sideband-resolved, such that $\omega_2 \gg \kappa$, and the auxiliary optical drive is strong enough to cool the auxiliary mechanical mode to the ground state. We later study a modified setup involving two optical cavity modes, in which neither the mechanical mode to be cooled nor the auxiliary mechanical mode need to be sideband-resolved.

The cooling scheme we present is a result of a destructive interference effect and is related to so-called EIT cooling, where electromagnetically induced transparency (EIT) \cite{Harris1997PhysToday} is exploited to suppress unwanted transitions when laser cooling atoms \cite{Morigi2000PRL,Roos2000PRL}. In particular, our proposals exploits OMIT, the optomechanical analogue of EIT, to modify the photon number fluctuation spectrum in the cavity and thus the relative rate of Stokes and anti-Stokes scattering. EIT cooling of the mechanical motion of larger objects has been proposed before, with setups including a superconducting qubit \cite{Xia2009PRL}, an atomic ensemble \cite{Genes2011PRA}, or an additional optical cavity \cite{Gu2013PRA,Liu2013CLEO}. However, the cooling scheme we present is to our knowledge the first to take advantage of OMIT, and it requires only optomechanical interactions. Finally, we note that cavity-assisted ground state cooling of mechanical motion in the unresolved sideband regime can also be achieved in systems exhibiting dissipative optomechanical coupling \cite{Elste2009PRL,Andre2011PRL,Weiss2013PRA,Weiss2013NJP}, where the Stokes processes are also suppressed by a destructive interference effect, or in standard optomechanical systems with time-dependent optical driving \cite{Wang2011PRL,Machnes2012PRL,Vanner2013NatComm}.
    
This article is organized as follows. We first revisit the theory of cavity-assisted sideband cooling of mechanical motion in Sec.~\ref{sec:Review}. This provides a background for our discussion in Sec.~\ref{sec:CoolingMethod1}, where we present a scheme for ground state cooling of a mechanical mode in the unresolved sideband regime. This scheme involves one cavity mode coupled to two mechanical modes, where the auxiliary mechanical mode is sideband-resolved. In Sec.~\ref{sec:CoolingMethod2}, we introduce a modified setup consisting of two cavity modes and two mechanical modes, where ground state cooling of one mode is possible even when both mechanical oscillators are in the unresolved sideband regime. We conclude in Sec.~\ref{sec:Conclusion}.

\section{Review of sideband cooling of mechanical motion}
\label{sec:Review}
We will first briefly review the theory of cavity-assisted sideband cooling of mechanical motion, following Refs.~\cite{Marquardt2007PRL,Wilson-Rae2007PRL}. We assume that the mechanical oscillator has resonance frequency $\omega_1$, and that its motion is described by the position operator $\hat{x}_1 = x_\mathrm{zpf,1} (\hat{b}_1 + \hat{b}_1^\dagger)$, where $\langle \hat{x}_1 \rangle = 0$ and $\hat{b}_1$ is the phonon annihilation operator. The constant $x_\mathrm{zpf,1}$ is the size of the zero point fluctuations and is given by $x_\mathrm{zpf,1} = \sqrt{\hbar/(2m_1\omega_1)}$, where $m_1$ is the effective oscillator mass. The optical cavity is coherently driven at a frequency $\omega_{d}$ which is detuned from the cavity resonance frequency $\omega_c$ by $\Delta = \omega_{d} - \omega_c$. The photon annihilation operator is $\hat{a}$ and the cavity linewidth will be denoted by $\kappa$.

The motion of the mechanical oscillator modulates the cavity resonance frequency according to the Hamiltonian
\begin{equation}
\label{eq:Hint}
H_\mathrm{int} = \hbar g_1 \left(\hat{b}_1 + \hat{b}_1^\dagger \right) \left(\hat{a}^\dagger \hat{a} - |\bar{a}|^2\right) ,
\end{equation}
where $g_1$ is a coupling rate. The constant $|\bar{a}|^2$ is the average cavity photon number, such that the last parenthesis is the photon number fluctuations. We define the photon number fluctuation spectrum as
\begin{equation}
\label{eq:PhotFluctSpect}
S_{nn}[\omega] = \int_{-\infty}^{\infty} d\tau e^{i\omega\tau}\overline{\langle\delta \hat{n}(t+\tau)\delta \hat{n}(t) \rangle}\end{equation}
where $\delta \hat{n} = \hat{a}^\dagger \hat{a} - |\bar{a}|^2$ and the bar indicates a time average. 

The photon number fluctuations give rise to a noisy force on the mechanical oscillator, which cause transitions between phonon Fock states of the mechanical oscillator. In a Fermi Golden Rule approach \cite{Clerk2010RMP,Marquardt2007PRL,Wilson-Rae2007PRL}, the rate for optically induced transitions from $n$ to $n+1$ phonons (Stokes scattering) is
\begin{equation}
\label{eq:GammaUp}
\Gamma_{n \rightarrow n+1} = g_1^2 (n+1) S_{nn}[-\omega_1] .
\end{equation}
This refers to the unperturbed photon number fluctuation spectrum $S_{nn}[\omega]$ in the absence of optomechanical interaction, i.e.~for $g_1 = 0$. Oppositely, the rate for optically induced transitions from $n+1$ to $n$ phonons (anti-Stokes scattering) is 
\begin{equation}
\label{eq:GammaDown}
\Gamma_{n+1 \rightarrow n}  = g_1^2 (n+1) S_{nn}[\omega_1] .
\end{equation}
The Fermi Golden Rule approach is valid as long as the single-photon coupling rate $g_1 \ll \kappa,\omega_1$, which is the case for almost all experiments to date. It also requires that the fluctuation spectrum $S_{nn}[\omega]$ changes slowly on the scale given by the effective linewidth of the mechanical oscillator, to be defined below.

In addition to the optically induced transition rates, we assume that the oscillator is connected to a thermal bath, which in the absence of optomechanical coupling gives rise to an intrinsic linewidth $\gamma_1$ and an average phonon occupation number $n_{\text{th,1}} = 1/(e^{\hbar\omega_1/k_B T} - 1) \approx k_B T/\hbar \omega_1$, where $T$ is the bath temperature.

The phonon Fock state occupation probalities $p_n$ can now be determined by solving the rate equation 
\begin{eqnarray}
\label{eq:RateEq}
\dot{p}_n & = & - \left[\gamma_1 n_\text{th,1} (n+1) p_n + \gamma_1 (n_\text{th,1} + 1) n \right] p_n \notag \\
& + &  \gamma_1 n_\text{th,1} n \, p_{n-1} + \gamma_1 (n_\text{th,1} + 1)(n + 1) p_{n+1} \notag \\
& - & \left(\Gamma_{n \rightarrow n+1} + \Gamma_{n \rightarrow n-1} \right) p_n \notag \\
& + & \Gamma_{n-1 \rightarrow n} \, p_{n-1} + \Gamma_{n+1 \rightarrow n} \, p_{n+1} 
,
\end{eqnarray} 
which includes both the transitions induced by the thermal bath and the ones induced by photon number fluctuations (i.e.~photon shot noise). The solution to Eq.~\eqref{eq:RateEq} is a thermal distribution $p_n = n_{1}^n/(n_{1} + 1)^{n+1}$ with an effective average phonon number
\begin{equation}
\label{eq:AvPhonNum}
n_1 = \frac{\gamma_1 n_{\text{th,1}}  + \gamma_\mathrm{opt,1} n_\text{opt,1}}{\tilde{\gamma}} .
\end{equation}
We have defined the effective mechanical linewidth $\tilde{\gamma}_1 = \gamma_1 + \gamma_\mathrm{opt,1}$ where the optical contribution is
\begin{equation}
\label{eq:GammaOpt}
\gamma_\mathrm{opt,1} = g_1^2 \left(S_{nn}[\omega_1] - S_{nn}[-\omega_1] \right) .
\end{equation}
The expression for the average phonon number \eqref{eq:AvPhonNum} can be viewed as a weighted sum of the thermal occupation number $n_{\text{th,1}}$ and
\begin{equation}
\label{eq:noptDef}
n_\mathrm{opt,1} = \frac{1}{S_{nn}[\omega_1]/S_{nn}[-\omega_1] - 1} , 
\end{equation}
which is a measure of the effective temperature of the photon shot noise. 

We see that two criteria must be fulfilled in order to obtain ground state cooling in the sense of $n_{1} \ll 1$. The first requirement is that the optical broadening of the linewidth be large enough that $\gamma_\mathrm{opt,1} \gg \gamma_1 n_\text{th,1}$, which means that the rate at which phonons are removed from the oscillator by anti-Stokes scattering is large compared to the rate at which they enter from the thermal bath. The second requirement is that $n_\mathrm{opt,1} \ll 1$, which is the same as saying that Stokes scattering processes do not heat the mechanical oscillator, as they are rare compared to the anti-Stokes processes. The latter requirement is fulfilled whenever the photon fluctuation spectrum fulfills $S_{nn}[\omega_1] \gg S_{nn}[-\omega_1]$.

Let us now look at the standard example of a single mechanical oscillator and a single drive. The unperturbed photon number fluctuation spectrum is then \cite{Marquardt2007PRL}
\begin{equation}
\label{eq:SStandard}
S_{nn}[\omega] = \frac{\kappa |\bar{a}|^2}{(\kappa/2)^2 + (\omega + \Delta)^2} .
\end{equation}
The optical contribution to the mechanical linewidth becomes
\begin{equation}
\label{eq:gammaoptStandard}
\gamma_\mathrm{opt,1} = -\frac{4 g_1^2 \kappa |\bar{a}|^2 \omega_1 \Delta  }{\left[(\kappa/2)^2 + (\omega_1 + \Delta)^2\right]\left[(\kappa/2)^2 + (\omega_1 - \Delta)^2\right]} , 
\end{equation}
which is proportional to the number of photons in the cavity $|\bar{a}|^2$ and thus proportional to the power of the coherent drive. This means that for laser detuning $\Delta < 0$, one can in principle increase $\gamma_\mathrm{opt,1}$ by increasing the power of the optical drive. 

\begin{figure}[t]
\includegraphics[width=0.9\columnwidth]{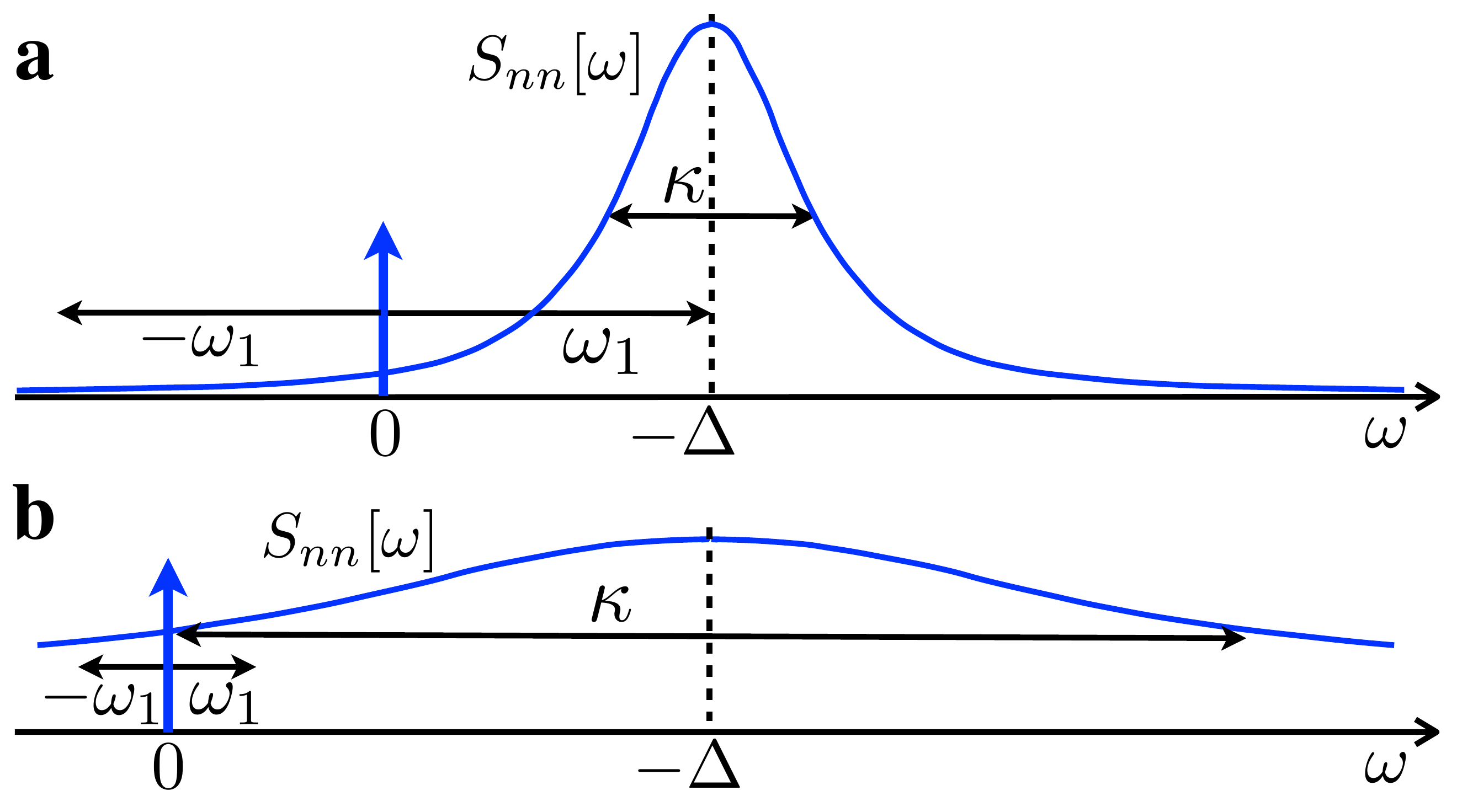}
\caption{The photon number fluctuation spectrum \eqref{eq:SStandard} for both the resolved (a) and unresolved sideband regime (b). Ground state cooling is only possible in the resolved sideband regime $\omega_1 > \kappa$, where $S[\omega_1] \gg S[-\omega_1]$.}
\label{fig:StandardcoolingFig}
\end{figure}

The number $n_\mathrm{opt,1}$ characterizing the photon shot noise does on the other hand not depend on laser power. The negative detuning that minimizes \eqref{eq:noptDef} is $\Delta = -\sqrt{(\kappa/2)^2 + \omega_1^2}$. In the resolved sideband limit $\omega_1 \gg \kappa$, the optimal detuning is thus $\Delta = -\omega_1$, which gives $n_\mathrm{opt,1} = \kappa^2/(4\omega_1)^2 \ll 1$. The small $n_\mathrm{opt,1}$ is a result of the large asymmetry between $S_{nn}[\omega_1]$ and $S_{nn}[-\omega_1]$, as depicted in Fig.~\ref{fig:StandardcoolingFig}.
However, in the unresolved sideband regime $\kappa \gg \omega_1$ with the detuning $\Delta = -\kappa/2$, we get $n_\mathrm{opt,1} = \kappa/(8\omega_1) \gg 1$. This corresponds to an effective temperature $T_\mathrm{eff} = \hbar \kappa/(8 k_B)$, which is usually referred to as the Doppler cooling limit. The large  $n_\mathrm{opt,1}$ is due to the lack of asymmetry between $S_{nn}[\omega_1]$ and $S_{nn}[-\omega_1]$, as shown in Fig.~\ref{fig:StandardcoolingFig}. This shows that even if you satisfy the first criterion for ground state cooling, $\gamma_\mathrm{opt,1} \gg \gamma_1 n_\text{th,1}$, the heating due to photon shot noise as described by the second term in Eq.~\eqref{eq:AvPhonNum} forbids ground state cooling in the unresolved sideband regime. In other words, Stokes scattering is only suppressed in the resolved sideband regime.

After having highlighted the obstacle to achieving ground state cooling in the unresolved sideband regime $\kappa \gg \omega_1$, we will now move on to a cooling scheme that circumvents this problem.

\section{Ground state cooling in the unresolved sideband regime with an auxiliary mechanical mode}
\label{sec:CoolingMethod1}

\subsection{Idea}

\begin{figure}[h]
\includegraphics[width=0.99\columnwidth]{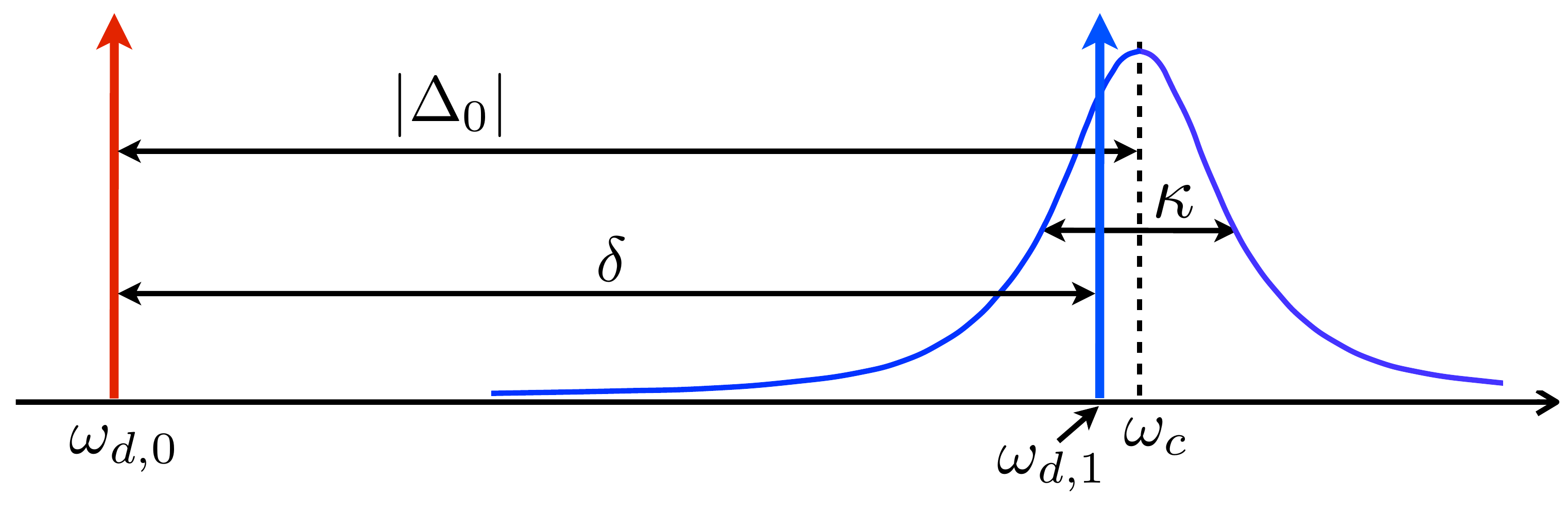}
\caption{Setup for the proposed cooling scheme. There are two optical drives, one at frequency $\omega_{d,0}$ (beam 0) and one at $\omega_{d,1}$ (beam 1). The off-resonant beam 0 and mechanical mode 2 will lead to squeezing of the photon number fluctuation spectrum, such that Stokes scattering from beam 1 is suppressed. This allows for ground state cooling of mechanical mode 1 even though it is in the unresolved sideband regime.}
\label{scheme}
\end{figure}

In this section, we will show that it is possible to cool a low-frequency mechanical oscillator to the ground state with the aid of an additional high-frequency mechanical oscillator and a second optical drive. The mechanical mode we wish to cool will be referred to as mode 1. Its resonance frequency is $\omega_1$ and we assume that $\omega_1 \ll \kappa$, where $\kappa$ is the optical cavity linewidth. The auxiliary mechanical mode is referred to as mode 2 and has resonance frequency $\omega_2$, and we will assume that this is large compared to the cavity linewidth, i.e.~$\omega_2 \gg \kappa$. The coherent drive that will cool mode 1 will be referred to as beam 1 and will have the frequency $\omega_{d,1}$, which is detuned from the cavity resonance frequency $\omega_c$ by $\Delta_1 = \omega_{d,1} - \omega_c = -\omega_1$. The idea is to add another laser drive, beam 0, at the frequency $\omega_{d,0}$, which is far detuned from the cavity resonance frequency by $\Delta_0 = \omega_{d,0} - \omega_\mathrm{c} = -\omega_2 - 2\omega_1$. Our setup is shown in Fig.~\ref{scheme}.

For sufficiently strong driving, the auxiliary beam 0 will lead to squeezing of the photon number fluctuations in the cavity. The squeezing results from destructive interference due to quantum optomechanical correlations between the motion of the auxiliary mechanical mode and the vacuum noise of the electromagnetic field - the same correlations that are responsible for the experimentally demonstrated ponderomotive squeezing in optomechanical systems \cite{Brooks2012Nature,Purdy2013PRX,Safavi-Naeini2013Nature}. With our choice of drive detunings, this interference can lead to a suppression of Stokes scattering from beam 1 due to mechanical mode 1, which is exactly what we need in order to be able to cool mode 1 to the quantum ground state.   

\subsection{Model}

We now present the model describing the setup for ground state cooling depicted in Fig.~\ref{scheme}. The Hamiltonian of the system is
\begin{equation}\label{eq1}
H=H_0+H_{\text{int,1}}+H_{\text{int,2}},
\end{equation}
where 
\begin{equation}\label{free}
H_0=\hbar\omega_c\hat{a}^\dag\hat{a}+\hbar\omega_1\hat{b}_1^\dag\hat{b}_1+\hbar\omega_2\hat{b}_2^\dag\hat{b}_2
\end{equation}
is the contribution from the optical cavity mode and the two mechanical modes when they are not coupled. The two mechanical modes can belong to physically separate systems, or they can be different normal modes of the same mechanical system. The position operator for mechanical mode $i$ is $\hat{x}_i = x_{\mathrm{zpf},i} (\hat{b}_i + \hat{b}_i^\dagger)$. Due to the presence of two drives in our scheme, we cannot {\it a priori} assume that $\langle\hat{x}_i\rangle = 0$ as in Sec.~\ref{sec:Review}, but we can and do assume that its time average is zero, i.e.~that is has no DC (or zero frequency) component. The creation and annihilation operators satisfy the bosonic commutation relations and the mechanical modes are well separated in frequency as discussed above. The coupling between the cavity and the mechanical modes is due to the radiation pressure interaction
\begin{equation}\label{coup}
H_{\text{int,i}}=\hbar g_i \left(\hat{b}_i+\hat{b}_i^\dag \right)\left(\hat{a}^\dag\hat{a}- \langle \hat{a}^\dagger \hat{a} \rangle_\text{DC} \right),
\end{equation}
where the constant $\langle \hat{a}^\dagger \hat{a} \rangle_\text{DC}$ is the DC component of the photon number expectation value and will be specified later. Note that the inclusion of the term $\langle \hat{a}^\dagger \hat{a} \rangle_\text{DC}$ in the interaction Hamiltonian is consistent with the assumption that $\langle \hat{x}_i \rangle$ has no DC component. In other words, the average displacement from the average radiation pressure has been taken into account when defining $\hat{x}_i$. Coherent driving and dissipation are accommodated by the input-output formalism \cite{Collett1984PRA,Clerk2010RMP} discussed below.

\subsection{Equations of motion }

Our scheme for cooling mechanical mode 1 relies on suppressing Stokes scattering from beam 1 to the frequency $\omega_c - 2\omega_1$. The suppression arises from the interaction between the cavity mode and mechanical mode 2 in presence of the external driving. We may thus calculate the photon number fluctuation spectrum in the absence of mechanical mode 1, and afterwards calculate the average phonon occupation number in mode 1 using the Fermi Golden Rule approach. Therefore, we proceed by solving the coupled dynamics of the cavity and mode 2, ignoring mode 1. The validity of this approach will be commented on in Sec.~\ref{GScooling}. In Sec.~\ref{sec:BeyondFGR}, we will also go beyond the Fermi Golden Rule approach and take the full dynamics of the three-mode problem into account.

We employ the input-output formalism \cite{Collett1984PRA,Clerk2010RMP} to include dissipation and optical driving. Moving to a frame rotating at $\omega_{d,0}$, i.e.~$\hat{a}(t) \rightarrow e^{-i \omega_{d,0} t} \hat{a}(t)$, the equations of motion for the cavity and mechanical mode 2 become
\begin{align}
\dot{\hat{a}} & =-\left(\frac{\kappa}{2}-i\Delta_0\right)\hat{a}-ig_2 \left(\hat{b}_2+\hat{b}_2^\dag \right)\hat{a}+\sqrt{\kappa}\hat{a}_{\text{in}} \label{eom1} \\
\dot{\hat{b}}_2 & =-\left(\frac{\gamma_2}{2}+i\omega_2\right)\hat{b}_2-ig_2\left(\hat{a}^\dag\hat{a}- \langle \hat{a}^\dagger \hat{a} \rangle_\text{DC}\right)+\sqrt{\gamma_2}\hat{\eta}_2 . \label{eom12}
\end{align} 
We have defined $\Delta_0=\omega_{d,0}-\omega_c = -\omega_2 - 2\omega_1$ and $\gamma_2$ is the intrinsic linewidth of mechanical mode 2. The input operator $\hat{a}_{\text{in}} = \bar{a}_{0,\text{in}} + e^{-i\delta t}\bar{a}_{1,\text{in}} +\hat{\xi}(t)$ contains the coherent input amplitudes $\bar{a}_{j,\text{in}}$ due to drives 0 and 1, as well as vacuum noise $\hat{\xi}(t)$. The frequency difference between the drives has been defined as $\delta = \omega_{d,1} - \omega_{d,0} = \omega_1 + \omega_2$ (see Fig.~\ref{scheme}). The operator $\hat{\eta}_2$ describes thermal noise acting on mechanical mode 2. The correlation properties of the noise operators will be presented below. 

The two drive tones will lead to a beat note at frequency $\delta$ in the cavity field intensity. This beat note will result in an oscillating force on the mechanical mode 2, setting it into coherent motion. Due to the nonlinear nature of the optomechanical interaction, this motion will in general give rise to optical coherence at frequencies other than the original drive frequencies. We can write the cavity field operator as
\begin{equation}
\label{eq:Coherences}
\hat{a}(t) = \sum_{j = -\infty}^\infty e^{-i j \delta t} \bar{a}_j  + \hat{d}(t) ,
\end{equation}
where $\bar{a}_j$ is the coherent amplitude at frequency $\omega_{d,0} + j \delta$ and $\langle\hat{d}(t)\rangle \equiv 0$. Furthermore, we write the phonon annihilation operator as
\begin{equation}
\label{eq:PhononCoherence}
\hat{b}_2(t) = \sum_{j > 0} e^{-i j \delta t} \bar{b}_{2,j} + \hat{c}_2(t) 
\end{equation}
with $\langle \hat{c}_2 (t) \rangle \equiv 0$. The coherent amplitudes $\bar{a}_j$ and $\bar{b}_{2,j}$ are determined by taking the expectation value of Eqs.~\eqref{eom1} and \eqref{eom12}. To be consistent with Eq.~\eqref{eq:PhononCoherence}, we define $\langle \hat{a}^\dagger \hat{a} \rangle_\text{DC} = \sum_j |\bar{a}_j|^2$ and keep only the positive frequency part of $\langle \hat{a}^\dagger \hat{a} \rangle$ in the equation for $\langle \hat{b}_2\rangle$. We also neglect small terms of the form $\langle \hat{c}_2 \hat{d} \rangle$ etc. This gives the nonlinear equations
\begin{eqnarray}
\label{eq:EOMClassic}
\left[\frac{\kappa}{2} - i(j\delta + \Delta_0)\right]\bar{a}_j & = & \sqrt{\kappa} \left(\bar{a}_\mathrm{in,0} \delta_{j,0} + \bar{a}_\mathrm{in,1} \delta_{j,1}\right) \quad  \\
& - & i g_2 \sum_{j' > 0} \left(\bar{b}_{2,j'} \bar{a}_{j-j'} + \bar{b}^\ast_{2,j'} \bar{a}_{j+j'} \right) \notag  \\
\left[\frac{\gamma_2}{2} - i(j\delta - \omega_2)\right] \bar{b}_{2,j} & = & -i g_2 \sum_{j'} \bar{a}^\ast_{j'} \bar{a}_{j+j'}  .
\end{eqnarray}
To first order in the optomechanical coupling $g_2$, these equations give
\begin{align}\label{class1}
\bar{a}_0 & =\frac{\sqrt{\kappa}\bar{a}_{0,\text{in}} }{\kappa/2-i\Delta_0},  \\ 
\bar{a}_1 & =\frac{\sqrt{\kappa}\bar{a}_{1,\text{in}} }{\kappa/2-i\Delta_1}, \nonumber \\
\bar{b}_{2,1} & = \frac{-ig_2\bar{a}_0^*\bar{a}_1}{\gamma_2/2-i (\delta -\omega_2) } , \nonumber
\end{align}
and zero for all other amplitudes. Here, $\bar{a}_0$ and $\bar{a}_1$ are just the cavity amplitudes in the absence of a mechanical oscillator, whereas $\bar{b}_{2,1}$ is the amplitude of the coherent motion of resonator 2 resulting from those amplitudes. We have introduced $\Delta_1 = \omega_{d,1} - \omega_c = -\omega_1$. To second order in $g_2$, one finds corrections to the amplitudes $\bar{a}_0$ and $\bar{a}_1$, as well as nonzero values for $\bar{a}_{-1}$ and $\bar{a}_2$. However, as long as we assume
\begin{equation}
\label{eq:AssumptionG}
\frac{|G_{2,0}|^2}{\kappa \omega_1} \ , \ \frac{|G_{2,1}|^2}{\omega_2 \omega_1} \ll 1 ,
\end{equation}
with the definition 
\begin{equation}
\label{eq:GijDef}
G_{i,j} = g_i \bar{a}_j , 
\end{equation}
these corrections are small compared to the empty cavity result and can be neglected. We may therefore proceed with the first order results in \eqref{class1}. 

Inserting the expressions \eqref{class1} in Eqs.~(\ref{eom1}) and (\ref{eom12}) gives the equations of motion for the fluctuations in the cavity mode and mechanical mode 2:
\begin{align}
\dot{\hat{d}} & =-\left(\frac{\kappa}{2}-i\Delta_0\right)\hat{d}-i(\hat{c}_2+\hat{c}_2^\dag)\left(G_{2,0}+G_{2,1} e^{-i\delta t}\right) \notag \\ 
& -\frac{i}{\omega_1}\left(G_{2,0}^*G_{2,1} e^{-i\delta t}+ G_{2,0}G_{2,1}^* e^{i\delta t}\right)\hat{d}+\sqrt{\kappa}\hat{\xi} \label{eom2}\\
\dot{\hat{c}}_2 & =-\left(\frac{\gamma_2}{2}+i\omega_2\right)\hat{c}_2-i\left(G_{2,0}^*+G_{2,1}^* e^{i\delta t}\right)\hat{d} \notag \\
& -i\left(G_{2,0}+G_{2,1} e^{-i\delta t}\right) \hat{d}^\dag+\sqrt{\gamma_2}\hat{\eta}_2,\label{eom22}
\end{align}
where we have ignored products of two fluctuation operators such as $\hat{d}^\dag\hat{d}$. By inspecting Fig.~\ref{scheme}, it is clear that it is only the anti-Stokes processes from beam 0 that are resonant. This allows for a simplification of the equations (\ref{eom2}) and (\ref{eom22}) where we only keep the resonant optomechanical interaction terms. Also, the off-resonant third term on the right-hand side of (\ref{eom2}), arising from the coherent motion caused by the beat note, can be neglected as we have implicitly assumed $|G_{2,1}||G_{2,0}|/(\omega_1\omega_2)\ll 1$. These considerations lead to the simplified equations  
\begin{align}
\dot{\hat{d}} & =-\left(\frac{\kappa}{2}-i\Delta_0\right)\hat{d}-iG_{2,0}\left(\hat{c}_2 + \hat{c}_2^\dagger\right)+\sqrt{\kappa}\hat{\xi} \label{eom3}\\
\dot{\hat{c}}_2 &=-\left(\frac{\gamma_2}{2}+i\omega_2\right)\hat{c}_2-i \left(G_{2,0}^*\hat{d} + G_{2,0} \hat{d}^\dagger\right)+\sqrt{\gamma_2}\hat{\eta}_2 .  \label{eom32}
\end{align}
Note that we could have simplified these equations even further by performing a rotating wave approximation. However, for later convenience, we keep the counter-rotating terms.

We now define the Fourier transform by $\hat{f}[\omega]=\int_{-\infty}^{\infty} dt \, e^{i\omega t} \hat{f}(t)$ and write their arguments in square brackets. We adopt a convention according to which the transform of the hermitian conjugate is defined by $\hat{f}^{\dag}[\omega]=\int_{-\infty}^{\infty} dt \, e^{i\omega t} \hat{f}^\dag(t)$, such that $\hat{f}^\dag[\omega]=(\hat{f}[-\omega])^\dag$. By Fourier transforming Eqs.~\eqref{eom3} and \eqref{eom32} we get linearly coupled algebraic equations. Defining the dimensionless position operator $\hat{z}_2 = \hat{c}_2 + \hat{c}_2^\dagger$, we find the solution
\begin{eqnarray}
\hat{z}_2[\omega] & = & \frac{1}{N[\omega]} \Big[ \sqrt{\gamma_2} \left(\chi_2^{-1 \, \ast}[-\omega] \hat{\eta}_2[\omega] +  \chi_2^{-1}[\omega] \hat{\eta}^\dagger_2[\omega] \right) \label{fluc} \\
& - & 2\sqrt{\kappa}\omega_2  \left(G_{2,0}^\ast \chi_{c}[\omega] \hat{\xi}[\omega] + G_{2,0} \chi^\ast_{c}[-\omega] \hat{\xi}^\dagger[\omega] \right) \Big] \nonumber \quad  \\
\hat{d}[\omega] & = & \chi_c[\omega] \left(\sqrt{\kappa} \hat{\xi}[\omega] - i G_{2,0} \hat{z}_2[\omega] \right).  \label{fluc2}
\end{eqnarray}
The fluctuation operators are expressed in terms of the susceptibilities
\begin{eqnarray}
\label{eq:Suscept}
\chi_c[\omega] & = & \frac{1}{\kappa/2-i(\omega+\Delta_0)} \\
\chi_2[\omega] & = & \frac{1}{\gamma_2/2 - i(\omega - \omega_2)},
\end{eqnarray}
and the quantity
\begin{equation}
\label{eq:Ndef}
N[\omega] = \chi_2^{-1}[\omega] \chi_2^{-1 \ast}[-\omega] - 2 i \omega_2 |G_{2,0}|^2 \left(\chi_{c}[\omega] - \chi_{c}^\ast[-\omega]\right) .
\end{equation}

We are now in a position to calculate any correlation function involving $\hat{c}_2$ and $\hat{d}$ from the correlation properties of the noise operators $\hat{\xi}$ and $\hat{\eta}_2$. The optical vacuum noise satisfies $\langle \hat{\xi}[\omega]\hat{\xi}^\dag[\omega'] \rangle=2\pi\delta(\omega+\omega')$ and $\langle \hat{\xi}^\dag[\omega]\hat{\xi}[\omega'] \rangle=0$, whereas the thermal noise from the mechanical bath obeys $\langle \hat{\eta}_2[\omega]\hat{\eta}_2^\dag[\omega'] \rangle=2\pi(n_{\text{th},2}+1 )\delta(\omega+\omega')$ and $\langle \hat{\eta}_2^\dag[\omega]\hat{\eta}_2[\omega'] \rangle=2\pi n_{\text{th},2}\delta(\omega+\omega')$. The number $n_{\text{th},2} = 1/(e^{\hbar\omega_2/k_B T} - 1) $ is the average phonon occupation number of mode 2 in the absence of coupling to the cavity.

\subsection{Photon number fluctuation spectrum} 
The interaction between the cavity and mechanical mode 1 is given by the Hamiltonian (\ref{coup}) with $i = 1$. Following the Fermi Golden Rule approach reviewed in Sec.~\ref{sec:Review}, the effective average phonon number and the effective linewidth for mode 1 is determined by the photon number fluctuation spectrum $S_{nn}[\omega]$ at frequencies $\pm \omega_1$. Having solved the coupled dynamics of the cavity and mechanical mode 2, we can now calculate the spectrum $S_{nn}[\omega]$.

The photon number fluctuation operator is in this case $\delta \hat{n} = \hat{a}^\dagger \hat{a} - (|\bar{a}_0|^2 + |\bar{a}_1|^2)$. By inserting this as well as the displacement decomposition $\hat{a}(t)=\bar{a}_0+e^{-i\delta t}\bar{a}_1+\hat{d}(t)$ into Eq.~(\ref{eq:PhotFluctSpect}), we obtain
\begin{align}\label{noise2}
S_{nn}[\omega] & = |\bar{a}_0|^2 \left(S_{\hat{d}\hat{d}^\dag}[\omega]+S_{\hat{d}^\dag\hat{d}}[\omega] \right) \\
& + |\bar{a}_1|^2 \left(S_{\hat{d}\hat{d}^\dag}[\omega+\delta]+S_{\hat{d}^\dag\hat{d}}[\omega-\delta] \right) \nonumber \\
& + 2 \pi |\bar{a}_0|^2|\bar{a}_1|^2 \left[\delta(\omega - \delta) + \delta(\omega + \delta) \right], \nonumber
\end{align}
where we have defined the spectra
\begin{align}
S_{\hat{d}\hat{d}^\dag}[\omega] & =\int_{-\infty}^{\infty} \frac{d\omega'}{2\pi}\langle\hat{d}[\omega]\hat{d}^{\dag}[\omega']\rangle, \label{noised} \\
S_{\hat{d}^\dag\hat{d}}[\omega] & =\int_{-\infty}^{\infty} \frac{d\omega'}{2\pi}\langle\hat{d}^{\dag}[\omega]\hat{d}[\omega']\rangle \label{noised2}
\end{align}
and dropped contributions from products of four fluctuation operators $\hat{d}, \hat{d}^{\dag}$ since those are negligible when $|\bar{a}_i|^2\gg1$.

When inserting Eq.~\eqref{noise2} into the expressions for the transition rates \eqref{eq:GammaUp} and \eqref{eq:GammaDown}, the first line in Eq.~\eqref{noise2} should be interpreted as giving rise to Stokes and anti-Stokes scattering from beam 0 to frequencies $\omega_{d,0} \pm \omega_1$, by creation and annihilation of phonons in mode 1. The second line gives rise to Stokes and anti-Stokes scattering from beam 1 to frequencies $\omega_{d,1} \pm \omega_1$. As is evident from Fig.~\ref{scheme}, the scattering processes from beam 0 due to the low-frequency mode 1 are suppressed relative to the ones from beam 1, since the frequency of beam 1 is well within the cavity susceptibility whereas beam 0 is far off resonance. The third line in Eq.~\eqref{noise2} is a result of the beat note between the two beams at frequency $\delta$, which can in principle induce coherent motion in mode 1. Formally, this term will not contribute to $S_{nn}[\pm \omega_1]$, since $\delta \neq \pm \omega_1$. In reality, however, it is conceivable that an off-resonant beat note can drive mode 1. This would not be captured by the simple Fermi Golden Rule approach we take here. Nevertheless, since the beat note frequency is so far detuned from the resonance frequency of mechanical mode 1 in this case, $\delta - \omega_1 = \omega_2 \gg \omega_1$, we will assume that mode 1 does not respond to the beat note driving at all. This point is further discussed in Sec.~\ref{sec:BeatNote}. 

By employing the expression for $\hat{d}[\omega]$ given in Eq.~(\ref{fluc2}), the spectra (\ref{noised}) and (\ref{noised2}) become
\begin{align}
S_{\hat{d}\hat{d}^\dag}[\omega] & = \kappa |\chi_c[\omega]|^2 \label{noised3}  \\
& \times \left(1 +\frac{|G_{2,0}|^2}{\kappa} S_{\hat{z}_2 \hat{z}_2}[\omega]   + \frac{2 G_{2,0}}{\sqrt{\kappa}} \mathrm{Im} \, S_{\hat{z}_2 \hat{\xi}^\dagger}[\omega] \right)  \nonumber \\
S_{\hat{d}^\dag\hat{d}}[\omega] & = |G_{2,0}|^2 |\chi_c[-\omega]|^2 S_{\hat{z}_2 \hat{z}_2}[\omega].\label{noised32}
\end{align}
The first term in the paranthesis in \eqref{noised3} is the empty cavity fluctuation spectrum originating from vacuum noise. The second term is expressed in terms of the mechanical position spectrum
\begin{eqnarray}
\label{eq:SzzDef}
S_{\hat{z}_2 \hat{z}_2}[\omega] & = & \int_{-\infty}^\infty \frac{d\omega'}{2\pi} \langle \hat{z}_2[\omega] \hat{z}_2[\omega'] \rangle , 
\end{eqnarray}
whereas the third term originates from the correlation between the oscillator position and the vacuum noise:
\begin{eqnarray}
\label{eq:SzxiDef}
S_{\hat{z}_2 \hat{\xi}^\dagger}[\omega] & = & \int_{-\infty}^\infty \frac{d\omega'}{2\pi} \langle \hat{z}_2[\omega] \hat{\xi}^\dagger[\omega'] \rangle . 
\end{eqnarray}
It is the term \eqref{eq:SzxiDef} that will lead to a suppression of the photon number fluctuation spectrum and thereby permit ground state cooling of mechanical mode 1.

Let us first examine the mechanical position spectrum, which in the limit $|G_{2,0}| \ll \kappa$ becomes
\begin{eqnarray}
\label{eq:SzzRes}
S_{\hat{z}_2 \hat{z}_2}[\omega] & = & \tilde{\gamma}_2\left[|\tilde{\chi}_2[\omega]|^2(\tilde{n}_2+1)+|\tilde{\chi}_2[-\omega]|^2\tilde{n}_2  \right].
\end{eqnarray}
We have introduced the effective susceptibility for mechanical mode 2
\begin{equation}
\label{eq:SusceptEff}
\tilde{\chi}_{2}[\omega] =  \frac{1}{\tilde{\gamma}_2/2-i(\omega-\tilde{\omega}_2)} , 
\end{equation}
with the effective linewidth and frequency:
\begin{eqnarray}
\tilde{\gamma}_2 & = & \gamma_2 + 2 |G_{2,0}|^2 \mathrm{Re} \, \left(\chi_c[\omega_2] - \chi_c^\ast[-\omega_2] \right) \label{eq:gamma2Eff} \\
\tilde{\omega}_2 & = & \omega_2 + |G_{2,0}|^2 \mathrm{Im} \, \left(\chi_c[\omega_2] - \chi_c^\ast[-\omega_2] \right) . \label{eq:omega2Eff}
\end{eqnarray}
Note that the spectrum \eqref{eq:SzzRes}, consisting of two Lorentzians at $\omega = \pm \tilde{\omega}_2$, is that of a harmonic oscillator in a thermal state, where the number $\tilde{n}_2$ is the average phonon number in mechanical mode 2 and is given by
\begin{equation}
\label{eq:n2Def}
\tilde{n}_2 = \langle \hat{c}^\dagger_2 \hat{c}_2 \rangle = \frac{\gamma_2 n_\text{th,2} + \kappa |G_{2,0}|^2 |\chi_c[-\omega_2]|^2}{\tilde{\gamma}_2} .
\end{equation}
In the setup we discuss here, we find that by using $|G_{2,0}|,\omega_1 \ll \kappa \ll \omega_2$, the effective parameters for mode 2 become $\tilde{\gamma}_2 \approx \gamma_2 + 4 |G_{2,0}|^2/\kappa$, $\tilde{\omega}_2 \approx \omega_2 - 8|G_{2,0}|^2 \omega_1/\kappa^2$, and $\tilde{n}_2 \approx \gamma_2 n_\text{th,2}/\tilde{\gamma}_2$. 

We now look at the interference term \eqref{eq:SzxiDef}. Inserting Eq.~\eqref{fluc} gives
\begin{equation}
\label{eq:SzxiRes}
S_{\hat{z}_2 \hat{\xi}^\dagger}[\omega] = - 2\sqrt{\kappa} \omega_2 G_{2,0}^\ast \frac{\chi_c[\omega] }{N[\omega]} . 
\end{equation}
We are interested in the spectrum $S_{nn}[\omega]$ at frequencies $\omega \sim \pm \omega_1$. This means that the spectra $S_{\hat{z}_2 \hat{z}_2}[\omega]$ and $S_{\hat{z}_2 \hat{\xi}^\dagger}[\omega]$, whose support is confined to frequencies $\sim \pm \tilde{\omega}_2$, will only contribute to the second line in Eq.~\eqref{noise2}. From this, it is clear that we only need the interference term at positive frequencies $\sim \tilde{\omega}_2$, which leads to the simplification
\begin{equation}
\label{eq:SzxiRes2}
S_{\hat{z}_2 \hat{\xi}^\dagger}[\omega] = - i \sqrt{\kappa} G_{2,0}^\ast \chi_c[\omega_2] \tilde{\chi}_2[\omega] \ , \quad \omega \sim \tilde{\omega}_2.
\end{equation}
Taking the imaginary value of this, as required by Eq.~\eqref{noised3}, gives two terms, one which is a Lorentzian in the frequency $\omega$ and one that has an anti-symmetric Fano-shape. 

We are now ready to write down an approximate expression for the photon number fluctuation spectrum $S_{nn}[\omega]$. At frequencies $\omega \sim -\omega_1$, it becomes
\begin{eqnarray}
\label{eq:SnnNegSec3}
& & S_{nn}[\omega] =  \frac{\kappa |\bar{a}_0|^2}{\omega_2^2}   \\
 & &  + \frac{4 |\bar{a}_1|^2 }{\kappa} \left\{1 + \frac{|G_{2,0}|^2 \left[\tilde{\gamma}_2 (\tilde{n}_2 - 1 ) - \frac{16 \omega_1}{\kappa} (\omega +\omega_1)  \right]}{\kappa \left[ (\tilde{\gamma}_2/2)^2 + (\omega + \omega_1)^2 \right] }  \right\} . \nonumber
\end{eqnarray}
We have used the fact that $\tilde{\omega}_2 - \omega_2 \ll \tilde{\gamma}_2$. The second term in the curly brackets is a Lorentzian with width $\tilde{\gamma}_2$, and we observe that it can have a negative prefactor if $\tilde{n}_2 < 1$. This means that the photon number fluctuations can be suppressed at the frequency $-\omega_1$. This suppression originates from the interference term \eqref{eq:SzxiRes2}. The antisymmetric last term in the curly brackets is in this case not important and indeed vanishes for $\omega = - \omega_1$. In fact, we could have avoided this term altogether by shifting the drive frequencies such that $\Delta_0 = -\omega_2$ and $\Delta_1 = +\omega_1$, but this would not have made any difference for the ability to cool mode 1. 

\begin{figure}
\includegraphics[width=0.99\columnwidth]{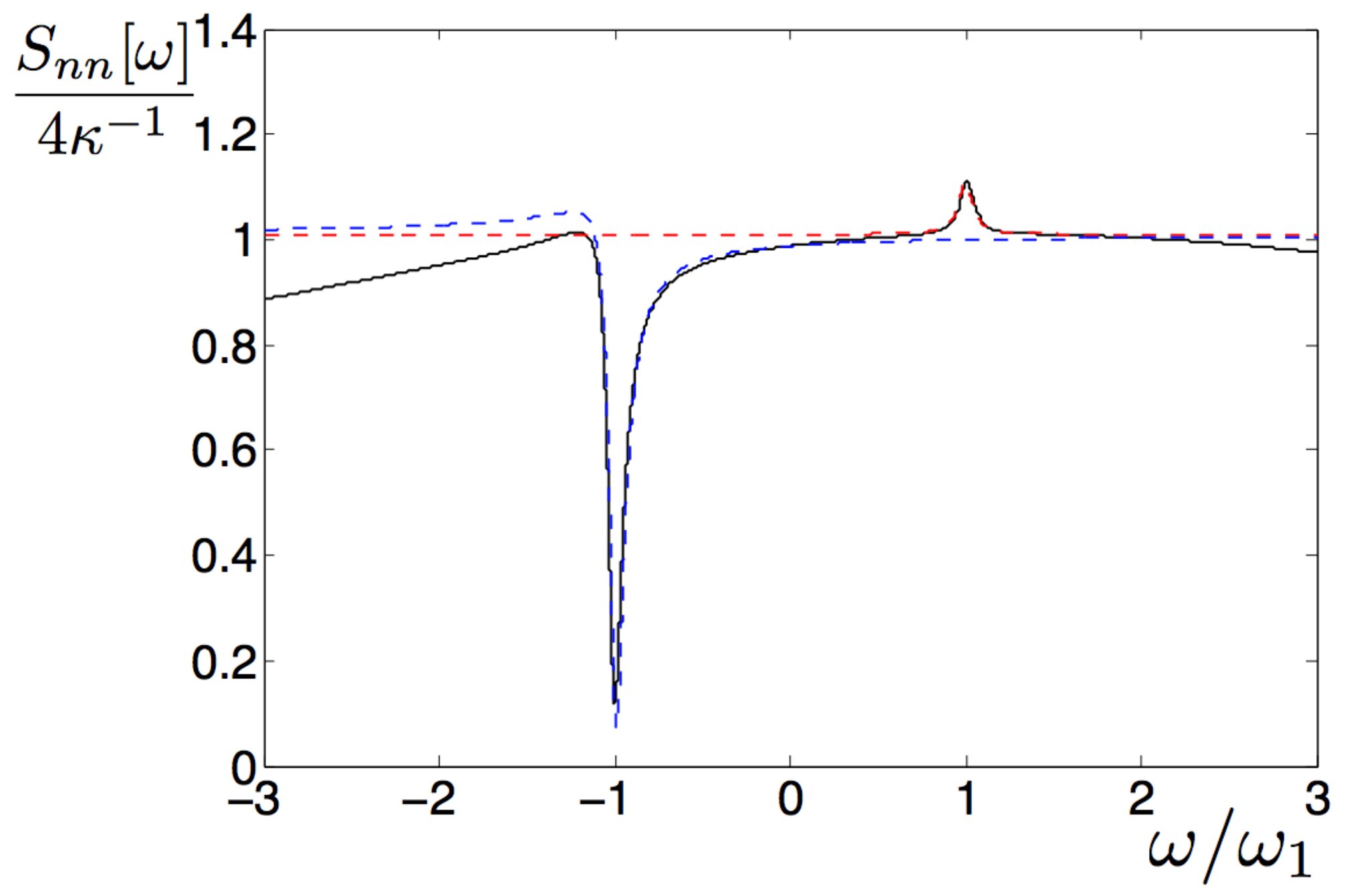}
\caption{The photon number fluctuation spectrum $S_{nn}[\omega]$ for the setup discussed here. The parameters are $\omega_1/\kappa = 0.05$, $\omega_2/\kappa = 5$, $\omega_2/\gamma_2 = 10^5$, $C_{2,0} = 100$ and $n_\text{th,2} = 10$. The presence of an auxiliary mechanical mode 2 and the additional drive (beam 0) leads to a dip in the spectrum at $\omega = -\omega_1$. This gives a large asymmetry between $S_{nn}[\omega_1]$ and $S_{nn}[-\omega_1]$ and ground state cooling of mechanical mode 1 is thus possible. The dashed curves are the approximate expressions \eqref{eq:SnnNegSec3} and \eqref{eq:SnnPosSec3}, and the solid line is the expression in the second line of \eqref{noise2}. We see that the approximations are accurate for $\omega \sim -\omega_1$ (Eq.~\eqref{eq:SnnNegSec3}) and for $\omega \sim \omega_1$ (Eq.~\eqref{eq:SnnPosSec3}).}
\label{SpectrSec3}
\end{figure}

The suppression of $S_{nn}[\omega]$ does not take place at positive frequencies $\omega \sim \omega_1$, where
\begin{equation}
\label{eq:SnnPosSec3}
S_{nn}[\omega] = \frac{\kappa |\bar{a}_0|^2}{\omega_2^2}  + \frac{4 |\bar{a}_1|^2 }{\kappa}\left(1 + \frac{|G_{2,0}|^2 \tilde{\gamma}_2 \tilde{n}_2/\kappa}{(\tilde{\gamma}_2/2)^2 + (\omega - \omega_1)^2}\right) .
\end{equation}
The second term in \eqref{eq:SnnPosSec3} is always positive and can only enhance the fluctuation spectrum. The spectrum $S_{nn}[\omega]$ is plotted in Fig.~\ref{SpectrSec3}, where a clear asymmetry between $S_{nn}[\omega_1]$ and $S_{nn}[-\omega_1]$ can be observed.

To quantify the asymmetry, let us now assume that the cooperativity 
\begin{equation}
\label{eq:CooperativitySec3}
C_{2,0} = 4 |G_{2,0}|^2/(\kappa\gamma_2) \gg 1 ,
\end{equation}
which means that the effective linewidth $\tilde{\gamma}_2 \gg \gamma_2$ and that $4|G_{2,0}|^2/(\kappa \tilde{\gamma}_2) \approx 1$. Inserting $\omega = \pm \omega_1$ in the above expressions then gives
\begin{eqnarray}
\label{eq:Snnpmomega1}
S_{nn}[-\omega_1] & = & \frac{\kappa |\bar{a}_0|^2}{\omega_2^2} + \frac{4 |\bar{a}_1|^2 }{\kappa} \tilde{n}_2 \\
S_{nn}[\omega_1] & = & \frac{\kappa |\bar{a}_0|^2}{\omega_2^2} + \frac{4 |\bar{a}_1|^2 }{\kappa} \left(\tilde{n}_2 + 1\right) . 
\end{eqnarray}
We see that to obtain a large asymmetry between positive and negative frequencies, two criteria must be satisfied. First of all, we need the average phonon number in mechanical mode 2 to be small, i.e.
\begin{equation}
\label{eq:n2Small}
\tilde{n}_2 \ll 1 . 
\end{equation}
In other words, the high-frequency mechanical mode 2 must either be cooled close to the ground state by beam 0, or have a vanishing thermal occupation number $n_\text{th,2}$. Secondly, we must require
\begin{equation}
\label{eq:Assumptiona0}
\frac{\kappa |\bar{a}_0|}{2\omega_2 |\bar{a}_1| } \ll 1.
\end{equation}
This ensures that the Stokes scattering from beam 0 is small relative to the anti-Stokes scattering from beam 1, and it can be satisfied for a mechanical mode 2 sufficiently far in the resolved sideband limit $\kappa \ll \omega_2$.

\subsection{Ground state cooling} \label{GScooling}

We are now equipped to calculate the average phonon number of mode 1. Using Eq.~\eqref{eq:noptDef} and the assumptions \eqref{eq:n2Small} and \eqref{eq:Assumptiona0}, we find
\begin{eqnarray}
\label{eq:nopt1Result}
n_\text{opt,1} & = & \tilde{n}_2 + \left(\frac{\kappa |\bar{a}_0|}{2\omega_2 |\bar{a}_1|}\right)^2 , 
\end{eqnarray}
which is the main result of this section. It is clear from Eqs.~\eqref{eq:AvPhonNum} and \eqref{eq:nopt1Result} that the average phonon number $n_1 \ll 1$ if all our above assumptions are satisfied and $\gamma_1 n_\text{th,1} \ll \tilde{\gamma}_1$. The effective linewidth of mode 1 becomes
\begin{equation}
\label{eq:LinewidthResult}
\tilde{\gamma}_1 \approx \frac{4 |G_{1,1}|^2}{\kappa}
\end{equation}
which follows from Eq.~\eqref{eq:GammaOpt}. The cooling scheme is effective as long as $\tilde{\gamma}_1$ is small compared to the width of the Lorentzian dip at $\omega = - \omega_1$ in the spectrum $S_{nn}[\omega]$, which is also necessary for our Fermi Golden Rule calculation to be valid. This means that we must assume $\tilde{\gamma}_1 \ll \tilde{\gamma}_2$, giving
\begin{equation}
\label{eq:AssumptionGammaTilde}
|G_{1,1}| \ll |G_{2,0}|
\end{equation}
as an additional requirement. See however Sec.~\ref{sec:BeyondFGR}, where we go beyond the Fermi Golden Rule approach.

It is worth reiterating that the ability to cool mode 1, which is {\it not} in the resolved sideband regime, is facilitated by the ability to cool mode 2, which {\it is} in the resolved sideband regime. This can be understood physically by realizing that the dip in the spectrum $S_{nn}[\omega]$ that prohibits Stokes scattering from beam 1 is a result of destructive interference in a similar way as in OMIT. This destructive interference occurs because a photon at $\omega_c - 2\omega_1$ can be down-converted to the drive frequency $\omega_{d,0}$ by creation of a phonon in mechanical mode 2, and subsequently up-converted again by destroying the phonon. The interference is however scrambled by thermal phonons in mode 2, which makes it necessary for beam 0 to be strong enough that it cools mode 2 to the ground state (or to have $n_\text{th,2} \ll 1$).

\subsection{Beyond Fermi Golden Rule}
\label{sec:BeyondFGR}

The result \eqref{eq:nopt1Result} is the minimal average phonon number achievable for the low-frequency oscillator. This limit is reached when $|G_{1,1}|$ is large enough that $\tilde{\gamma}_1 \gg \gamma_1 n_\text{th,1}$. However, one might not be able to reach this limit before violating the criterion \eqref{eq:AssumptionGammaTilde}, at which point the Fermi Golden Rule approach is no longer valid. In that case, the true minimal phonon number must be determined from a non-perturbative calculation. We will now outline this calculation and derive the average phonon number without the limitation \eqref{eq:AssumptionGammaTilde}.  

We return to the equations of motion for the fluctuation operators, but we now include the low-frequency oscillator 1 as well. As before, we assume that oscillator 1 is not affected by the beat note, such that $\hat{b}_1 \approx \hat{c}_1$ where $\langle \hat{c}_1 \rangle = 0$. When ignoring off-resonant terms as before, the equations of motion become
\begin{eqnarray}
\dot{\hat{d}} & = & -\left(\frac{\kappa}{2}-i\Delta_0\right)\hat{d} + \sqrt{\kappa}\hat{\xi} \label{eom3exact} \\
& - & iG_{2,0}\left(\hat{c}_2 + \hat{c}_2^\dagger\right) - i G_{1,1} e^{-i \delta t} \left(\hat{c}_1 + \hat{c}_1^\dagger \right)   \notag \\
\dot{\hat{c}}_2 & = & -\left(\frac{\gamma_2}{2}+i\omega_2\right)\hat{c}_2 + \sqrt{\gamma_2}\hat{\eta}_2  \label{eom32exact} \\
 & - & i \left(G_{2,0}^*\hat{d} + G_{2,0} \hat{d}^\dagger\right)  \notag  \\
\dot{\hat{c}}_1 & = & -\left(\frac{\gamma_1}{2}+i\omega_1\right)\hat{c}_1 + \sqrt{\gamma_1}\hat{\eta}_1 \label{eom33exact}  \\
& - & i \left(G_{1,1}^* e^{i \delta t} \hat{d} + G_{1,1} e^{-i \delta t} \hat{d}^\dagger\right)  \notag .
\end{eqnarray}
Here, the noise operator $\hat{\eta}_1$ fulfills the same correlation properties as $\hat{\eta}_2$, but with $n_\text{th,2} \rightarrow n_\text{th,1}$. Since the cavity decay rate $\kappa$ will be much larger than the effective decay rates of the mechanical oscillators, Eq.~\eqref{eom3exact} gives the approximate expression
\begin{eqnarray}
\label{eq:dExact}
\hat{d}(t) & = & \zeta(t) - i G_{2,0} \left(\chi_c[\omega_2] \hat{c}_2(t) + \chi_c[-\omega_2] \hat{c}_2^\dagger(t) \right) \\
& - & i G_{1,1} e^{-i \delta t} \left(\chi_c[\omega_1 + \delta] \hat{c}_1(t) + \chi_c[-\omega_1 + \delta] \hat{c}_1^\dagger(t) \right) \notag
\end{eqnarray}
for the cavity field fluctuations. We have defined
\begin{equation}
\label{eq:zetadef}
\zeta(t) = \sqrt{\kappa} \int_{-\infty}^t d \tau \, e^{-(\kappa/2 - i \Delta_{0})(t -\tau)} \hat{\xi}(\tau).
\end{equation}
Inserting \eqref{eq:dExact} into Eqs.~\eqref{eom32exact} and \eqref{eom33exact} gives coupled equations for the operators $\hat{c}_1$ and $\hat{c}_2$. We can neglect off-resonant terms, which results in the simplified equations
\begin{eqnarray}
\dot{\hat{c}}_2 & = & - \left(\frac{\tilde{\gamma}_2}{2} + i \tilde{\omega}_2\right)\hat{c}_2 + \sqrt{\gamma_2} \hat{\eta}_2 - i G_{2,0} \zeta^\dagger \ \label{eq:EOMc1c2} \\
& - & i G_{2,0}^\ast \left(\zeta - i G_{1,1} e^{-i\delta t} \chi_c[-\omega_1 + \delta] \hat{c}_1^\dagger  \right)  \notag \\
\dot{\hat{c}}_1 & = & - \left(\frac{\bar{\gamma}_1}{2} + i \bar{\omega}_1\right)\hat{c}_1 + \sqrt{\gamma_1} \hat{\eta}_1 - i G_{1,1}^\ast e^{i \delta t} \zeta  \label{eq:EOMc1c22} \\
& - & i G_{1,1}  e^{-i\delta t} \left(\zeta^\dagger + i G_{2,0}^\ast \chi_c^\ast[\omega_2] \hat{c}_2^\dagger \right)  \notag .
\end{eqnarray}
The parameters $\tilde{\gamma}_2$ and $\tilde{\omega}_2$ are defined as before (Eqs.~\eqref{eq:gamma2Eff} and \eqref{eq:omega2Eff}). We have also introduced the parameters
\begin{eqnarray}
\label{eq:bars}
\bar{\gamma}_1 & = & \gamma_1 + 2 |G_{1,1}|^2 \mathrm{Re} \, \left(\chi_c[\omega_1 + \delta] - \chi_c^\ast[-\omega_1 + \delta] \right) \ \\
\bar{\omega}_1 & = & \omega_1 + |G_{1,1}|^2 \mathrm{Im} \, \left(\chi_c[\omega_1 + \delta] - \chi_c^\ast[-\omega_1 + \delta] \right)
\end{eqnarray}
which would be the effective linewidth and frequency for oscillator 1 in the absence of beam 0, i.e.~when $G_{2,0} = 0$.

The coupled equations \eqref{eq:EOMc1c2} and \eqref{eq:EOMc1c22} can easily be solved in Fourier space, which gives
\begin{eqnarray}
\label{eq:c1Solution}
\hat{c}_1[\omega] & = & \Big\{ \sqrt{\gamma_1} \hat{\eta}_1[\omega] \\
& - & i\left(G_{1,1}^\ast \zeta[\omega + \delta] + G_{1,1} \zeta^\dagger[\omega - \delta] \right) \notag  \\
& + & G_{1,1} G_{2,0}^\ast \chi_c^\ast[\omega_2] \tilde{\chi}_2^\ast[-\omega + \delta] \Big[\sqrt{\gamma_2} \hat{\eta}_2^\dagger[\omega - \delta] \notag \\
&  + & i \left(G_{2,0}^\ast \zeta[\omega - \delta] + G_{2,0} \zeta^\dagger[\omega - \delta] \right) \Big] \Big\} \notag \\
& \times & \Big(\frac{\bar{\gamma}_1}{2} - i(\omega - \bar{\omega}_1) \notag \\
& + & |G_{1,1}|^2 |G_{2,0}|^2 \chi_c^\ast[\omega_2] \chi_c^\ast[-\omega_1 + \delta] \tilde{\chi}_2^\ast[-\omega + \delta] \Big)^{-1} \notag 
\end{eqnarray}
for the fluctuations of the low-frequency oscillator 1. From this, we find that the average phonon number is
\begin{eqnarray}
\label{eq:PhononNumberExact}
n_1 & = & \int_{-\infty}^\infty \frac{d \omega}{2 \pi} \int_{-\infty}^\infty \frac{d \omega'}{2 \pi} \langle \hat{c}^\dagger[\omega] \hat{c}[\omega'] \rangle \\
& = & \int_{-\infty}^\infty \frac{d \omega}{2 \pi} \Big[ \gamma_1 n_\text{th,1} \notag \\
& + & \kappa |G_{1,1}|^2 |\chi_c[-\omega_1 + \delta]|^2 \Big|1 - |G_{2,0}|^2 \chi_c[\omega_2] \tilde{\chi}_2[\omega + \delta] \Big|^2 \notag \\
& + & |G_{1,1}|^2 |G_{2,0}|^2 |\chi_c[\omega_2]|^2 |\tilde{\chi}_2[\omega + \delta]|^2 \gamma_2 \left(n_\text{th,2} + 1\right) \Big] \notag \\
& \times & \Big| \frac{\bar{\gamma}_1}{2} - i (\omega + \bar{\omega}_1) \notag \\
& + & |G_{1,1}|^2 |G_{2,0}|^2 \chi_c[\omega_2] \chi_c[-\omega_1 + \delta] \tilde{\chi}_2[\omega + \delta] \Big|^{-2} . \notag
\end{eqnarray}
The frequency integral can be performed by contour integration in the complex plane. 

\begin{figure}
\includegraphics[width=0.99\columnwidth]{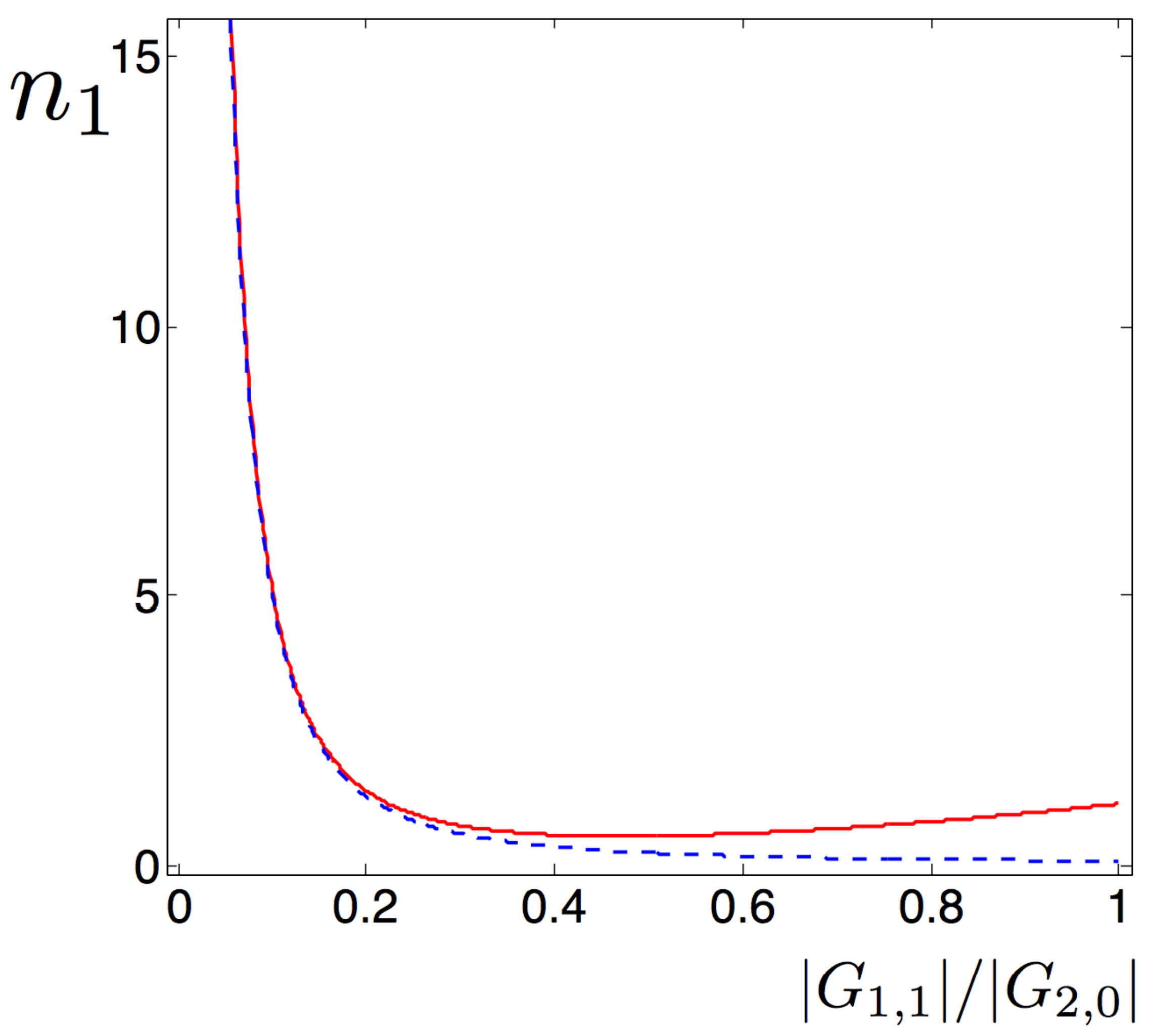}
\caption{The average phonon number $n_1$ of the low-frequency mechanical oscillator as a function of the coupling rate $|G_{1,1}|$. The solid line is the exact result in Eq.~\eqref{eq:n1Integrated} and the dashed curve is the result \eqref{eq:LimitFGR} from the Fermi Golden Rule approach. The thermal phonon number is $n_\text{th,1} = 500$, and the intrinsic mechanical $Q$ is $\omega_1/\gamma_1 = 10^5$. The other parameters are the same as in Fig.~\ref{SpectrSec3}. We observe that the two curves agree well for $|G_{1,1}| \ll |G_{2,0}|$, but for larger $|G_{1,1}|$ the exact result reaches a minimum given by Eqs.~\eqref{eq:optimal} and \eqref{eq:minn1}.}
\label{fig:ExactSec3}
\end{figure}

We now insert $\delta = \omega_1 + \omega_2$ and simplify the above expression by exploiting that $\chi_c[\omega_2] \approx 2/\kappa$ since $\kappa \gg \omega_1$. Furthermore, we assume $\tilde{\gamma}_1 \approx 4|G_{1,1}|^2/\kappa \gg \gamma_1$ and that $\bar{\gamma}_1 \approx \gamma_1 + 32 \tilde{\gamma}_1 (\omega_1/\kappa)^2 \ll \tilde{\gamma}_2$. The contour integration then gives the average phonon number
\begin{equation}
\label{eq:n1Integrated}
n_1 = \frac{\gamma_1 n_\text{th,1}}{\tilde{\gamma}_1} + \frac{ \gamma_1 n_\text{th,1} + \gamma_2 n_\text{th,2}}{\tilde{\gamma}_2} + \frac{\tilde{\gamma}_1 + \gamma_2}{\tilde{\gamma}_2} .
\end{equation}
In the limit $C_{2,0} \gg 1$ and $\tilde{\gamma}_2 \gg \tilde{\gamma}_1$, this reduces to
\begin{equation}
\label{eq:LimitFGR}
n_1 = \frac{\gamma_1 n_\text{th,1}}{\tilde{\gamma}_1} + \frac{\gamma_2 n_\text{th,2}}{\tilde{\gamma}_2} ,
\end{equation}
which is the result we got from the Fermi Golden Rule approach \footnote{The last term in \eqref{eq:nopt1Result} does not appear, since we have neglected off-resonant terms proportional to $G_{1,0}$.}. In Fig.~\ref{fig:ExactSec3}, we have plotted the exact result \eqref{eq:n1Integrated} together with the Fermi Golden Rule result \eqref{eq:LimitFGR}. We see that for small $|G_{1,1}|$, the two results agree, but that they start to deviate when $|G_{1,1}|$ becomes comparable to $|G_{2,0}|$. The exact curve reaches a minimum phonon number for a finite $|G_{1,1}|$.

With the result \eqref{eq:n1Integrated}, we can now determine the optimal value $\tilde{\gamma}_\text{1,min}$ that minimizes the average phonon number for oscillator 1. From $|G_\text{1,1,min}| = \sqrt{\kappa \tilde{\gamma}_\text{1,min}}/2$, one can then determine the optimal power of the cooling beam (beam 1). Minimization of $n_1$ gives
\begin{equation}
\label{eq:optimal}
\tilde{\gamma}_\text{1,min} = \sqrt{\gamma_1 n_\text{th,1} \tilde{\gamma}_2}
\end{equation}
and the minimum value in the limit $C_{2,0} \gg 1$ is
\begin{equation}
\label{eq:minn1}
n_\text{1,min} = \frac{\gamma_1 n_\text{th,1} +  \gamma_2 n_\text{th,2}}{\tilde{\gamma}_2} + 2\sqrt{\frac{\gamma_1 n_\text{th,1}}{\tilde{\gamma}_2}} .
\end{equation}
We see that when $\gamma_1 n_\text{th,1} \ll \tilde{\gamma}_2$, the average phonon number of oscillator 1 is limited by the phonon number of oscillator 2, as predicted by the Fermi Golden Rule approach. Finally, we point out that in the special case where the two oscillators have the same intrinsic $Q$ and their reservoirs have the same temperature $T$, we have $\gamma_1 n_\text{th,1} = \gamma_2 n_\text{th,2} = k_B T/(\hbar Q)$. For $\gamma_1 n_\text{th,1} \ll \tilde{\gamma}_2$, the result \eqref{eq:minn1} then simplifies to $n_1 = 2 \sqrt{\gamma_1 n_\text{th,1}/\tilde{\gamma}_2}$.

\subsection{The beat note}
\label{sec:BeatNote}

As mentioned above, the beat note in the optical intensity at frequency $\delta = \omega_1 + \omega_2$ can in principle induce coherent motion in mechanical mode 1 and thus prohibit ground state cooling. Similar to how we estimated the coherent amplitude in mode 2, $\bar{b}_{2,1}$, the estimate for the amplitude in mode 1 in units of zero point motion would be
\begin{equation}
\label{eq:bbar11}
|\bar{b}_{1,1}| \approx \frac{g_1}{\omega_2} |\bar{a}_0| |\bar{a}_1|.
\end{equation}
If we for simplicity assume that the two mechanical modes have the same intrinsic quality factor, i.e.~$Q_1 = Q_2$ with $Q_i = \omega_i/\gamma_i$, and that $n_\text{th,2} \gg 1$, the assumptions above gives
\begin{equation}
\label{eq:bbar112}
|\bar{b}_{1,1}| \gg \frac{\kappa n_\text{th,2}}{4 g_2 Q_2}
\end{equation}
For the coherent motion to be negligible, we would need $|\bar{b}_{1,1}| \ll 1$, and \eqref{eq:bbar112} thus puts a lower limit on the single-photon coupling strength $g_2$.

It is however important to note that the estimates \eqref{eq:bbar11} and \eqref{eq:bbar112} are not very realistic. The reason is that the drive is so far off resonance ($\delta - \omega_1 = \omega_2 \gg \omega_1$) that it is not reasonable to assume that mode 1 has a simple linear response at the drive frequency. In a real mechanical system, there are presumably many other normal modes between the resonance frequency $\omega_1$ and the beat note frequency. Thus, it is more conceivable that the beat note will drive other modes that are closer to it in frequency. Of course, one could then imagine that this would influence mode 1 indirectly through nonlinear interactions between the normal modes. All in all, the actual effect on mode 1 resulting from the beat note cannot be determined from the simple models we use here, but is likely to depend on which physical realization one is dealing with. We do however find the assumption to neglect its effect quite reasonable, as the drive frequency is so far off resonance. 

Finally, let us also mention that one could in principle cancel the effect of the beat note by adding more drives, and thus avoid the problem entirely. For example, if the optical cavity has another resonance at a very different frequency that also couples to the mechanical modes, one could add two additional drives at the second cavity resonance, separated by frequency $\delta$ in exactly the same way as beam 0 and 1. One could then adjust the powers and phases of the two new drives in such a way that the radiation pressure force from their beat note exactly cancels the one from the original beat note. One could also imagine adding more drives in the original cavity mode to remove the beat note, along the lines of Ref.~\cite{Steinke2013PRA}.

\subsection{Physical realizations}

The cooling scheme we have presented in this section relies on a separation of frequency scales between mechanical modes 1 and 2. One system where such a scheme could be of interest is the optical fiber cavity realization of the membrane-in-the-middle system \cite{Flowers-Jacobs2013ApplPhysLett}. These are systems where a thin dielectric membrane of typical width $d \sim 150$~nm is placed between two tiny fiber mirrors. Compared to the large scale version of the membrane-in-the-middle system \cite{Thompson2008Nature}, the optomechanical coupling strength is enhanced due to the short length ($\sim 80 \, \mu$m) of the fiber cavity. The short length of the cavity does however lead to a large cavity linewidth $\kappa \sim 100$~MHz, which is much larger than the typical resonance frequencies of the flexural mechanical modes of the membrane ($\sim 5$~MHz), such that the system resides in the unresolved sideband regime. 

The membrane does however not only have flexural modes, but also dilational modes \cite{Borkje2012NJP}. The fundamental dilational mode is simply an oscillation in the thickness of the membrane, which for a typical thickness of $d = 150$~nm has a frequency of $30$~GHz, well into the resolved sideband regime. The optomechanical coupling to such a dilational mode has to our knowledge not been experimentally studied, but it has been estimated that it can be significant provided the mechanical $Q$ of this mode is sufficiently large \cite{Borkje2012NJP}. One could thus envisage that the dilational mode could be used as the high-frequency mode 2 to enable ground state cooling of the lower-frequency flexural modes of the membrane. The large frequency of the dilational mode will also ensure a small thermal occupation number $n_\text{th,2}$ in mode 2, which is advantageous for our cooling scheme. For a dilational mode with frequency $\omega_2 = 2 \pi \times 30$~GHz at a temperature of $200$~mK, we have $n_\text{th,2} = 7 \times 10^{-4}$. For a square SiN$_3$ membrane with thickness $d = 150$~nm and sidelength $250 \ \mu$m, and a laser drive with wavelength $\sim 1 \ \mu$m and power $1.5$~mW, an estimate \cite{Borkje2012NJP} for the cooperativity is $C_{2,0} \sim 10^{-4} Q_2$, where $Q_2 = \omega_2/\gamma_2$ is the intrinsic quality factor of the dilational mode. This means that such a dilational mode could be used as the high-frequency mode in our cooling scheme if $Q_2 \gtrsim 10^5$. Note, however, that the mode would be broadened to an effective linewidth $\tilde{\gamma}_2 \approx \gamma_2 C_{2,0} \sim 10^{-4} \, \omega_2 = 2 \pi \times 3$~MHz. The frequency $\omega_1$ of the flexural mode to be cooled would need to be larger than $\tilde{\gamma}_2$ for our scheme to work. This frequency limit could potentially be lowered by using a thicker membrane with a smaller $\omega_2$. 

Our scheme is not only useful in systems where the cavity linewidth is large, but also in cases where one wishes to cool low frequency mechanical resonators. One example could be hybrid membrane-in-the-middle optomechanical systems, where the membrane motion also couples to an electrical circuit, facilitating conversion of micro- or radiowaves to optical light and vice versa \cite{Bagci2013,Andrews2013}. The membranes in such systems might need to be of a certain size in the transverse dimensions in order to accomodate coupling to both the optical cavity and to the electrical circuit, which can make reaching the well-resolved sideband regime challenging. Our scheme would still allow efficient conversion using a large area membrane by for example including an additional small area membrane to act as a high-frequency mode 2. The cooling scheme could potentially also be useful in hybrid systems where light couples to both membrane motion and to the collective motion of ultracold atoms \cite{Camerer2011PRL}, where it could allow for cooling of the low-frequency motion of the atomic ensemble.

\section{Ground state cooling with both mechanical modes in the unresolved sideband regime}
\label{sec:CoolingMethod2}

\subsection{Idea}

The cooling procedure presented in Sec.~\ref{sec:CoolingMethod1} requires an auxiliary sideband-resolved mechanical mode. Additionally, the beat note arising from the driving of the optical mode at two different frequencies could potentially limit the applicability of this cooling scheme (see discussion in Sec.~\ref{sec:BeatNote}). We now present a modified ground state cooling scheme which involves two optical modes and two mechanical modes. In this scheme, both mechanical modes can be in the unresolved sideband regime and the optical beat note does not drive the mechanical mode one wishes to cool. The idea is similar to that in 
Sec.~\ref{sec:CoolingMethod1}, in that one of the mechanical modes is used to squeeze the photon number fluctuations in one of the optical modes, thereby allowing an asymmetry in the rates of Stokes and anti-Stokes processes involving the other mechanical mode.

\subsection{Model}

We again consider a Hamiltonian of the form $H = H_0 + H_\text{int,1} + H_\text{int,2}$, where the term
\begin{align}\label{hh}
& H_0 = \hbar\omega_{c,0}\hat{a}_0^\dagger\hat{a}_0+\hbar\omega_{c,1}\hat{a}_1^\dagger\hat{a}_1+\hbar\omega_{1}\hat{b}_1^\dagger\hat{b}_1+\hbar\omega_{2}\hat{b}_2^\dagger\hat{b}_2 
\end{align}
describes the uncoupled optical ($\hat{a}_0,\hat{a}_1$) and mechanical modes ($\hat{b}_1,\hat{b}_2$). The optomechanical interaction for mechanical mode 1 is described by
\begin{align}\label{hh2}
H_\text{int,1} = \hbar g_1 (\hat{b}_1^\dagger+\hat{b}_1) & \Big[r \, \left(\hat{a}_0^\dagger\hat{a}_0 - \langle \hat{a}^\dagger_0 \hat{a}_0 \rangle_\text{DC}\right) \\
& + \, \left(\hat{a}_1^\dagger\hat{a}_1 - \langle \hat{a}^\dagger_1 \hat{a}_1 \rangle_\text{DC}\right) \Big] \notag ,
\end{align}
where $g_1$ ($r g_1$) is the coupling rate to optical mode 1 (0), and the constants $\langle \hat{a}^\dagger_j \hat{a}_j \rangle_\text{DC}$, which will be specified later, ensures that $\langle \hat{x}_1 \rangle$ has no DC component. The interaction with mechanical mode 2 is given by
\begin{align}\label{hh3}
& H_\text{int,2} = \hbar g_{2} (\hat{b}_2^\dagger+\hat{b}_2)\left(\hat{a}_0^\dagger\hat{a}_1+\hat{a}_1^\dagger\hat{a}_0 \right),
\end{align}
which is in an intercavity coupling Hamiltonian. 

The Hamiltonian $H$ arises naturally in systems where both the optical modes and the mechanical modes can have different symmetry properties. In general, the optomechanical coupling terms involving a mechanical mode $\hat{b}_i$ are of the form $H_{\text{int},i} = \sum_{j,k} \hbar g_{i,jk}(\hat{b}_i + \hat{b}_i^\dagger) \hat{a}^\dagger_j \hat{a}_k$, where $\hat{a}_j$ and $\hat{a}_k$ are annihilation operators for different cavity modes (or the same mode if $j = k$). The coupling rates $g_{i,jk}$ are typically overlap integrals of the mode functions. If the overall Hamiltonian is symmetric with respect to a particular operation (e.g.~inversion), the coupling rate $g_{i,jk}$ can only be nonzero if zero or two of the mode functions associated with $\hat{z}_i$, $\hat{a}^\dagger_j$, and $\hat{a}_k$ are antisymmetric with respect to the same operation. The interaction Hamiltonians Eqs.~\eqref{hh2} and \eqref{hh3} come about if for example the modes $\hat{b}_1$ and $\hat{a}_0$ are symmetric and the other modes antisymmetric with respect to an operation that leaves the physical system unchanged. See Sec.~\ref{sec:PhysRelSec4} for further details and potential realizations of this model.

\subsection{Equations of motion}

\begin{figure}
\includegraphics[width=0.99\columnwidth]{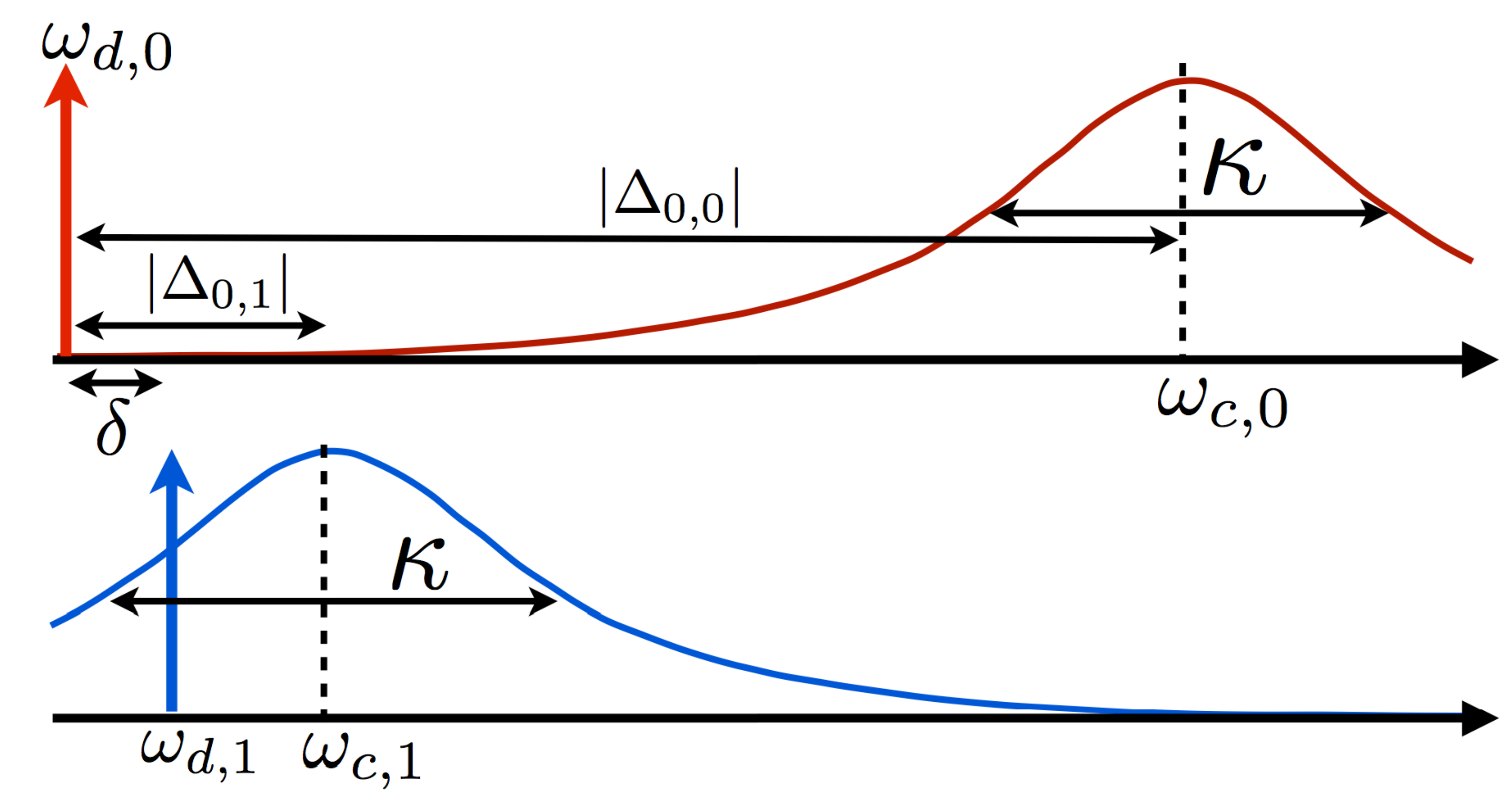}
\caption{Setup for the modified cooling scheme. We now have two optical cavity modes, 0 and 1, with resonance frequencies $\omega_{c,0}$ and $\omega_{c,1}$. Since mechanical mode 2 couples the two cavity modes, the drive in cavity mode 0 at frequency $\omega_{d,0}$ will squeeze the photon number fluctuation spectrum in cavity mode 1. This enables the drive in cavity mode 1 at $\omega_{d,1}$ to cool mechanical mode 1 to the quantum ground state. }
\label{SetupSec4}
\end{figure}

As depicted in Fig.~\ref{SetupSec4}, we are considering a situation where the optical modes 0 and 1 are subjected to external driving at frequencies $\omega_{d,0}$ and $\omega_{d,1}$, respectively. The role of beam 0 is to produce a desired photon number fluctuation spectrum in cavity mode 1, while beam 1 plays the role of the cooling laser of mechanical mode 1. In the following we are working in the frame rotating at the frequency $\omega_{d,0}$, so it is convenient to define the detunings $\Delta_{0,0}=\omega_{d,0}-\omega_{c,0} < 0$, $\Delta_{0,1}=\omega_{d,0}-\omega_{c,1} < 0$, and $\delta=\omega_{d,1}-\omega_{d,0} > 0$. $\Delta_{0,j}$ is the detuning of the drive frequency $\omega_{d,0}$ from the resonance frequency of cavity mode $j$, whereas $\delta$ is the frequency difference between beam 1 and 0. We assume $\omega_1,\omega_2,\delta \ll \kappa$ and will later see that ground state cooling of mechanical mode 1 becomes possible for a particular choice of $\delta$.

We now follow a similar approach as in Sec.~\ref{sec:CoolingMethod1} in that  we first solve the coupled dynamics of the cavity modes and mechanical mode 2 and then describe the cooling of mode 1 using the Fermi Golden Rule approach. Ignoring the optomechanical coupling to mechanical mode 1 and including dissipation in the standard way \cite{Collett1984PRA,Clerk2010RMP} gives the equations of motion in the frame rotating at $\omega_{d,0}$: 
 \begin{align}\label{eqs1}
\dot{\hat{a}}_0 & =-\left(\frac{\kappa}{2}-i\Delta_{0,0}\right)\hat{a}_0-ig_{2}(\hat{b}_2+\hat{b}_2^\dag)\hat{a}_1 \\
&+\sqrt{\kappa}\left(\bar{a}_{0,\text{in}}+\hat{\xi} _{0}\right)\nonumber\\
\dot{\hat{a}}_1 & =-\left(\frac{\kappa}{2}-i\Delta_{0,1}\right)\hat{a}_1-ig_{2}(\hat{b}_2+\hat{b}_2^\dag)\hat{a}_0 \nonumber \\
& +\sqrt{\kappa}\left( e^{-i\delta t}\bar{a}_{1,\text{in}}+\hat{\xi} _{1}\right)\nonumber\\
\dot{\hat{b}}_2 & =-\left(\frac{\gamma_2}{2}+i\omega_2\right)\hat{b}_2-ig_{2}\left(\hat{a}_0^\dag\hat{a}_1+\hat{a}_1^\dag\hat{a}_0\right)+\sqrt{\gamma_2}\hat{\eta}_2. \nonumber
\end{align}
The operators $\hat{\xi}_j$ describe vacuum noise entering cavity mode $j$, and $\hat{\eta}_2$ describes thermal noise from the mechanical bath as before.
For simplicity, we have assumed the cavity linewidths to be equal, since a difference between these will not influence our results in any significant way. As in Sec.~\ref{sec:CoolingMethod1}, we apply the approximate displacement transformations $\hat{a}_0=\bar{a}_0+\hat{d}_0(t)$, $\hat{a}_1=\bar{a}_1e^{-i\delta t}+\hat{d}_1(t)$ and $\hat{b}_2=\bar{b}_{2,1}e^{-i\delta t}+\hat{c}_2(t)$. The amplitudes are the same as in Eqs.~\eqref{class1} with the substitutions $\Delta_0 \rightarrow \Delta_{0,0}$ and $\Delta_1 \rightarrow \Delta_{0,1} + \delta$, and $\langle \hat{d}_0(t) \rangle = \langle \hat{d}_1(t) \rangle = \langle \hat{c}_2 (t) \rangle = 0$. This approximation is valid when $G^2_{2,j}/(\kappa\omega_1)\ll 1$, where $G_{2,j}=g_2\bar{a}_j$ as before.   

The equations of motion for the fluctuation operators $\hat{d}_0$, $\hat{d}_1$ and $\hat{c}_2$ can now be found by inserting the displacement transformations into Eqs.~\eqref{eqs1}. This gives quite complicated equations. However, for the cooling scheme to work, we need to assume that beam 0 is far off resonance from cavity mode 0, i.e.~$|\Delta_{0,0}| \gg \kappa$ (see Fig.~\ref{SetupSec4}). This means that the fluctuations $\hat{d}_0$ in cavity mode 0 are strongly suppressed at the relevant frequencies, such that Stokes scattering from beam 0 due to mechanical mode 1 is negligible. With this assumption, we can ignore the fluctuations $\hat{d}_0$ entirely and arrive at simpler equations of motion. The relevant dynamics is then described by 
\begin{align}\label{eqm4}
\dot{\hat{d}}_1 & =-\left(\frac{\kappa}{2}-i\Delta_{0,1}\right)\hat{d}_1-iG_{2,0}\left(\hat{c}_2+\hat{c}_2^\dagger\right)+\sqrt{\kappa}\hat{\xi}_{1}\\
\dot{\hat{c}}_2 & =-\left(\frac{\gamma_2}{2}+i\omega_2\right)\hat{c}_2-i\left(G_{2,0}^\ast \hat{d}_1+ G_{2,0}\hat{d}_1^\dagger \right) +\sqrt{\gamma_2} \hat{\eta}_{2}.
\end{align}  
These equations are identical to Eqs.~\eqref{eom3},\eqref{eom32} when substituting $\hat{d} \rightarrow \hat{d}_1$, $\hat{\xi} \rightarrow \hat{\xi}_1$ and $\Delta_0 \rightarrow \Delta_{0,1}$. The solution is given by Eqs.\eqref{fluc},\eqref{fluc2} subjected to the same substitutions. 

\subsection{Photon number fluctuation spectrum}

We can now determine the properties of mechanical mode 1 by again using the Fermi Golden Rule approach of Sec.~\ref{sec:Review}. This requires calculating the photon number fluctuation spectrum $S_{nn}[\omega]$ in Eq.~\eqref{eq:PhotFluctSpect} with the fluctuation operator now given by $\delta \hat{n} = r (\hat{a}^\dagger_0 \hat{a}_0 - \langle \hat{a}_0^\dagger \hat{a}_0\rangle_\text{DC}) + \hat{a}_1^\dagger \hat{a}_1 - \langle \hat{a}_1^\dagger \hat{a}_1\rangle_\text{DC}$. When calculating $S_{nn}[\omega]$, we again neglect the contributions containing four fluctuation operators. Additionally, we neglect the terms containing $\hat{d}_0$, in accordance with the earlier assumption that $|\Delta_{0,0}| \gg \kappa$. More precisely, we assume
\begin{equation}
\label{eq:NeglectMode0}
\frac{r |\bar{a}_0| \sqrt{(\kappa/2)^2 + \Delta_{0,1}^2}}{ |\bar{a}_1| |\Delta_{0,0}|} \ll 1 , 
\end{equation}
which ensures that the rate of anti-Stokes scattering from beam 1 is much larger than that of Stokes scattering from beam 0.

With $\langle \hat{a}_j^\dagger \hat{a}_j\rangle_\text{DC} = |\bar{a}_j|^2$, the photon number fluctuation spectrum becomes
\begin{equation}
\label{eq:SnnSec4Def}
S_{nn}[\omega] = |\bar{a}_1|^2 \left(S_{\hat{d}_1 \hat{d}_1^\dagger}[\omega + \delta] + S_{\hat{d}_1^\dagger \hat{d}_1}[\omega - \delta] \right)
\end{equation}
where the two terms in the paranthesis are defined as in Eqs.~\eqref{noised} and \eqref{noised2}. These terms are again given by Eqs.~\eqref{noised3} and \eqref{noised32}, and the spectra $S_{\hat{z}_2 \hat{z}_2}[\omega]$ and $S_{\hat{z}_2 \hat{\xi}_1^\dagger}[\omega]$  are still given by \eqref{eq:SzzRes} and \eqref{eq:SzxiRes} with the substitution $\Delta_{0} \rightarrow \Delta_{0,1}$. The effective linewidth, frequency, and average phonon number of mechanical mode 2 are also as defined in Sec.~\ref{sec:CoolingMethod1} (Eqs.~\eqref{eq:gamma2Eff},\eqref{eq:omega2Eff},\eqref{eq:n2Def}), but their values are now different since mechanical mode 2 is not sideband-resolved. We now find that the linewidth is
\begin{equation}
\label{eq:gammaTilde2Sec4}
\tilde{\gamma}_2 = \gamma_2 - \frac{4|G_{2,0}|^2 \kappa \Delta_{0,1} \omega_2 }{\left[(\kappa/2)^2 + \Delta_{0,1}^2\right]^2}  ,
\end{equation}
the resonance frequency is
\begin{equation}
\label{eq:omegaTilde2Sec4}
\tilde{\omega}_2 = \omega_2 + \frac{2|G_{2,0}|^2 \Delta_{0,1}}{\left[(\kappa/2)^2 + \Delta_{0,1}^2 \right]} ,
\end{equation}
and the effective average phonon number is
\begin{equation}
\label{eq:nTilde2Sec4}
\tilde{n}_2 = \frac{\gamma_2 n_\text{th,2}}{\tilde{\gamma}_2} + \frac{\kappa |G_{2,0}|^2}{\tilde{\gamma}_2 \left[(\kappa/2)^2 + \Delta_{0,1}^2 \right]} . 
\end{equation}
In arriving at these expressions, we have used the fact that $\omega_2 \ll \kappa$. The last term in Eq.~\eqref{eq:nTilde2Sec4} describes the contribution to the phonon occupation from radiation pressure shot noise (Stokes scattering). For $\tilde{\gamma}_2 \gg \gamma_2$, this term results in $\tilde{n}_2 \gg 1$, hindering ground state cooling of mechanical mode 2. However, we will see below that even though mechanical mode 2 will not be cooled to the ground state, its interaction with the optical field will facilitate ground state cooling of mechanical mode 1.

\begin{figure}
\includegraphics[width=0.99\columnwidth]{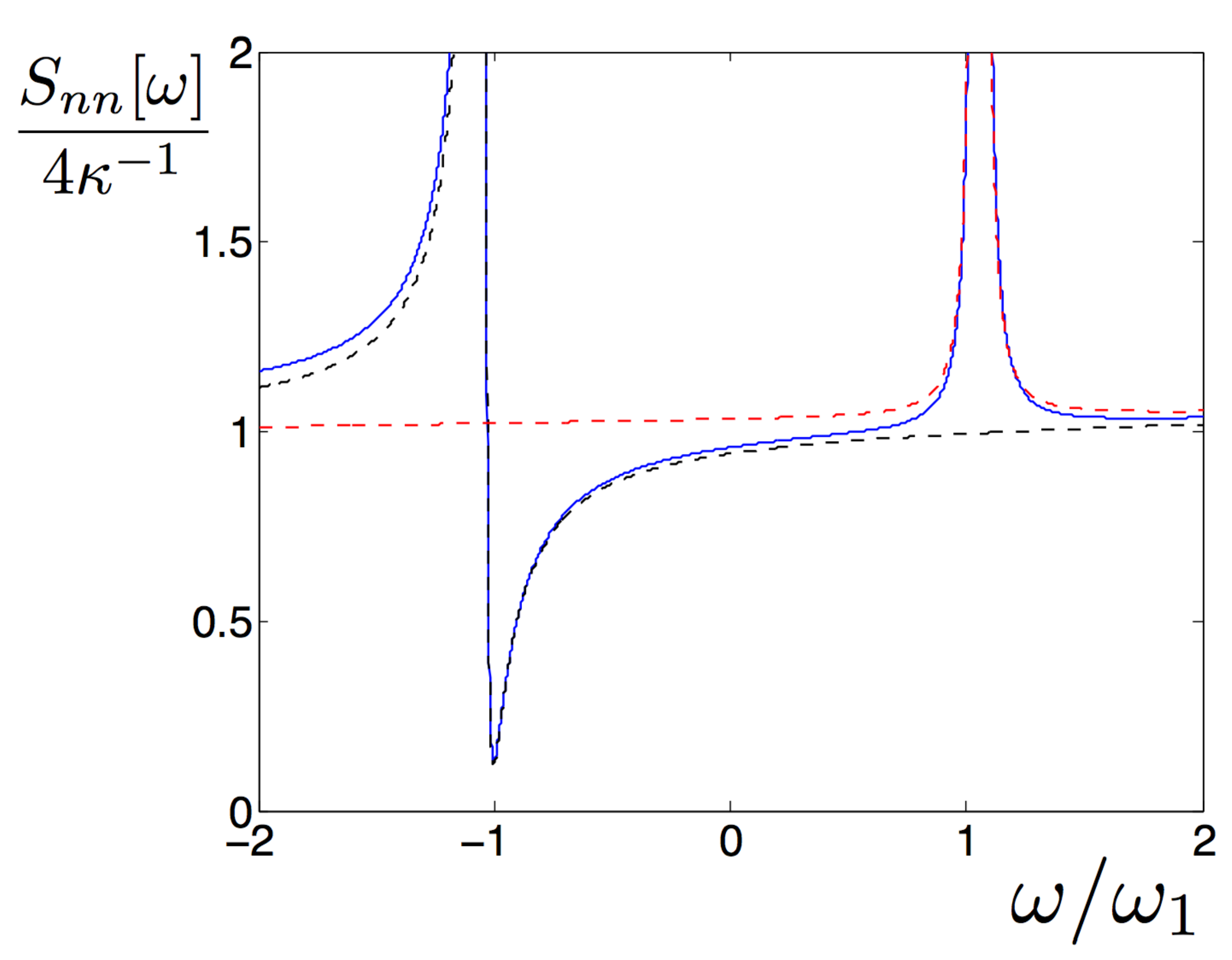}
\caption{The photon number fluctuation spectrum $S_{nn}[\omega]$ for the modified setup. The parameters are $\omega_1/\kappa = 0.01$, $\omega_2/\kappa = 0.02$, $\omega_2/\gamma_2 = 10^5$, $\Delta_{0,1}/\kappa = -2$, $C_{2,0} = 2 \times 10^4$ and $n_\text{th,2} = 100$. The optomechanical interaction with mechanical mode 2 leads to a dip at $\omega = -\omega_1$, enabling ground state cooling of mechanical mode 1. The dashed curves are the approximate expressions \eqref{eq:SNegFreq} and \eqref{eq:SPosFreq}, and the solid line is the expression in \eqref{eq:SnnSec4Def}.}
\label{SpectrSec4}
\end{figure}

We are now ready to write down approximate expressions for the photon number fluctuation spectrum. Let us define $u = \omega + \delta - \tilde{\omega}_2$ and assume that the detuning $|\Delta_{0,1}|$ is on the order of $\kappa$, such that $|\Delta_{0,1}| \gg \omega_1, \omega_2$. The spectrum for $|u| \ll \omega_1$ then becomes
\begin{eqnarray}
\label{eq:SNegFreq}
& & S_{nn}[\omega] = \frac{ \kappa |\bar{a}_1|^2}{ (\kappa/2)^2 + \Delta_{0,1}^2}  \\ & & \ \times \Bigg[1 + \frac{|G_{2,0}|^2}{(\tilde{\gamma}_2/2)^2 + u^2} \left(\frac{\tilde{\gamma}_2 \tilde{n}_2}{\kappa} + \frac{2 \Delta_{0,1} u}{ (\kappa/2)^2 + \Delta_{0,1}^2}\right) \nonumber \Bigg] .
\end{eqnarray}
We have assumed that $\tilde{n}_2 \gg 1$. The spectrum has a Lorentzian peak at $u = 0$, but also contains an asymmetric feature which comes from the interference term $S_{\hat{z}_2 \hat{\xi}^\dagger}[\omega]$ in Eq.~\eqref{eq:SzxiRes2}. Indeed, we see that for $u > 0$, the last term can lead to a suppression of the spectrum (since $\Delta_{0,1} < 0$). The spectrum has a minimum for $u = u_{\mathrm{min}}$ with
\begin{equation}
\label{eq:umin}
u_{\mathrm{min}} \approx - \frac{|G_{2,0}|^2}{\Delta_{0,1}} \left(1 + \epsilon\right),
\end{equation}
where 
\begin{equation}
\label{eq:epsilon}
\epsilon = \frac{4\left((\kappa/2)^2 + \Delta_{0,1}^2\right) n_\text{th,2}}{\kappa^2 C_{2,0}} ,
\end{equation}
$C_{2,0} = 4|G_{2,0}|^2/(\kappa \gamma_2)$ as before, and we have assumed $C_{2,0} \gg 1$. Thus, if we choose the frequency difference between the two drives to be 
\begin{equation}
\label{eq:deltaDef}
\delta = \omega_1 + \tilde{\omega}_2 + u_{\mathrm{min}},
\end{equation}
we find 
\begin{equation}
\label{eq:SNegFreq2}
S_{nn}[-\omega_1] \approx \frac{ \kappa |\bar{a}_1|^2}{ (\kappa/2)^2 + \Delta_{0,1}^2} \left(1 - \frac{\Delta_{0,1}^2(1 + \epsilon)^{-1}}{(\kappa/2)^2 + \Delta_{0,1}^2}\right) .
\end{equation}
We see that the spectrum is reduced relative to the empty cavity result for small values of $\epsilon$. See Fig.~\ref{SpectrSec4} for a plot of the spectrum $S_{nn}[\omega]$. The parameter $\epsilon$ is the ratio between the first and the second term in Eq.~\eqref{eq:nTilde2Sec4}, which are the contributions to the average phonon number from thermal fluctuations and from radiation pressure shot noise, respectively. In other words, small $\epsilon \ll 1$ means that the noise in the motion of mechanical mode 2 is predominantly due to radiation pressure shot noise. Note that the criterion $\varepsilon \ll 1$ is tantamount to $C_{2,0} \gg n_\text{th,2}$, which is the same criterion as we encountered in Sec.~\ref{sec:CoolingMethod1}. 

The spectrum at positive frequencies $\omega \sim \delta - \tilde{\omega}_2$ does not have the asymmetric feature resulting from the optomechanical correlations, and is simply a Lorentzian,
\begin{eqnarray}
\label{eq:SPosFreq}
S_{nn}[\omega] & = & \frac{ \kappa |\bar{a}_1|^2}{ (\kappa/2)^2 + \Delta_{0,1}^2}  \left(1 + \frac{\tilde{\gamma}_2 \tilde{n}_2 |G_{2,0}|^2}{\kappa\left[(\tilde{\gamma}_2/2)^2 + w^2\right]}  \right) , \quad 
\end{eqnarray}
with $w = \omega - \delta + \tilde{\omega}_2$. Inserting $\delta$ from Eq.~\eqref{eq:deltaDef} gives
\begin{equation}
\label{eq:SPosFreq2}
S_{nn}[\omega_1] \approx \frac{ \kappa |\bar{a}_1|^2}{ (\kappa/2)^2 + \Delta_{0,1}^2} \left(1 + \frac{\Delta_{0,1}^2(1 + \epsilon)^{-1}}{(\kappa/2)^2 + \Delta_{0,1}^2}\right) . 
\end{equation}
In the limit $\epsilon \ll 1$, the ratio between the spectrum at positive and negative frequencies becomes
\begin{equation}
\label{eq:RatioSpectrum}
\frac{S_{nn}[\omega_1]}{S_{nn}[-\omega_1]} = 1 + 8 \left(\frac{\Delta_{0,1}}{\kappa}\right)^2 . 
\end{equation}
We see that for a sufficiently large $|G_{2,0}|$ and $|\Delta_{0,1}| \sim \kappa$, there is a significant asymmetry between the spectrum at positive and negative frequencies, which is the prerequisite for cooling.


\subsection{Ground state cooling}

From Eqs.~\eqref{eq:noptDef}, \eqref{eq:SNegFreq2}, and \eqref{eq:SPosFreq2}, it follows that the limit on the average phonon number in mechanical mode 1 imposed by radiation pressure shot noise from beam 1 is
\begin{equation}
\label{eq:nOpt1Method2}
n_\text{opt,1} = \frac{(1 + \varepsilon)}{8} \left(\frac{\kappa}{\Delta_{0,1}}\right)^2 + \frac{\varepsilon}{2} \approx \frac{1}{8} \left(\frac{\kappa}{\Delta_{0,1}}\right)^2 ,
\end{equation}
where the last approximation assumes $\epsilon \ll 1$.
This is the main result of this section. Choosing for example $\Delta_{0,1} = -2\kappa$ gives $n_\text{opt,1} = 1/32 \ll 1$. Ultimately, one would of course also need to consider the limit imposed by Stokes scattering from beam 0 in cavity mode 0, which we neglected based on the assumption \eqref{eq:NeglectMode0}.

The Fermi Golden Rule approach we have used here is valid as long as the quantity $u_\mathrm{min}$ far exceeds the effective linewidth of mechanical mode 1, $\tilde{\gamma}_1 \approx \gamma_\text{1,opt}$, where
\begin{equation}
\label{eq:TildeGamma}
\gamma_\text{1,opt} = \frac{2 |G_{1,1}|^2 \kappa \Delta_{0,1}^2}{\left[(\kappa/2)^2 + \Delta_{0,1}^2\right]^2}  
\end{equation}
when $\varepsilon \ll 1$. This means that the Fermi Golden Rule approach requires
\begin{equation}
\label{eq:G11Limit}
|G_{1,1}| \ll  \frac{(\kappa/2)^2 + \Delta_{0,1}^2}{\sqrt{\kappa |\Delta_{0,1}|} |\Delta_{0,1}|} |G_{2,0}| \sim |G_{2,0}|.
\end{equation}
However, we will see in Sec.~\ref{sec:BeyondFGRSec4} that ground state cooling is possible and the Fermi Golden Rule result is fairly accurate for values of $|G_{1,1}|$ beyond this regime. This is convenient, since the coupling $|G_{1,1}|$, which is controlled by the power of beam 1, must be made large enough that $\gamma_1 n_\text{th,1} \ll \gamma_\text{1,opt}$.

Finally, we point out that for a system where $r$ can be made to vanish, one would not need the criterion $|\Delta_{0,0}| \gg \kappa$. This would allow for a cooling scheme where both drive frequencies could be close to a cavity resonance, and thereby reduce the amount of power needed in beam 0.

\subsection{Beyond Fermi Golden Rule}
\label{sec:BeyondFGRSec4}

To go beyond the regime described by Eq.~\eqref{eq:G11Limit}, we can again perform a non-perturbative calculation in the same way as in Sec.~\ref{sec:BeyondFGR}. In fact, the expression for the average phonon number given in Eq.~\eqref{eq:PhononNumberExact} is valid also here, when replacing $\Delta_0$ with $\Delta_{0,1}$. It is straightforward to check that the Fermi Golden Rule result is reproduced when inserting the approximations $\tilde{\chi}_2[\omega + \delta] \approx i/u$ and $\chi_c[\omega_2] \approx 1/(\kappa/2 - i\Delta_{0,1})$ into Eq.~\eqref{eq:PhononNumberExact}. 

\begin{figure}
\includegraphics[width=0.99\columnwidth]{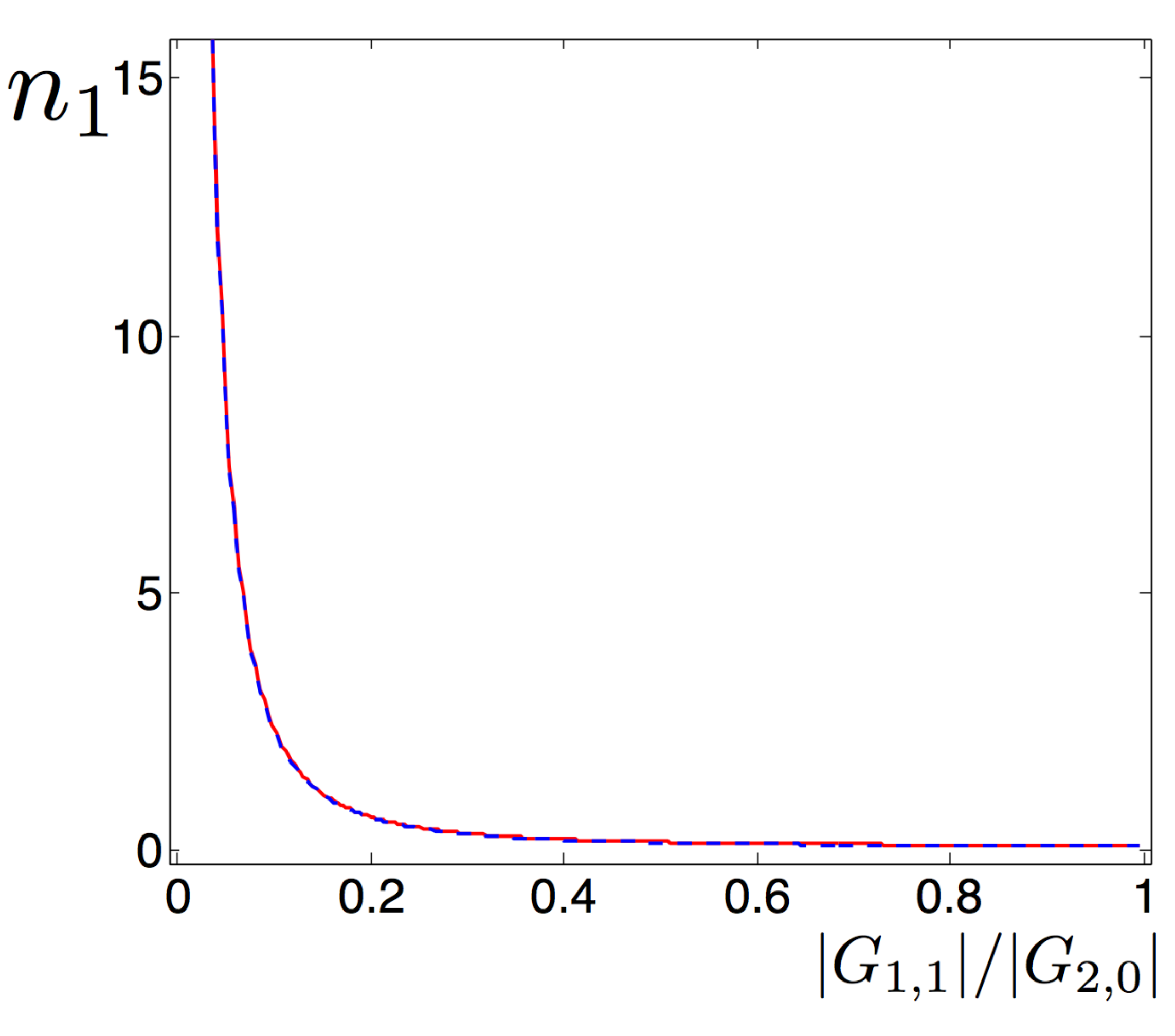}
\caption{The average phonon number $n_1$ of the low-frequency mechanical oscillator as a function of the coupling rate $|G_{1,1}|$. The exact result in Eq.~\eqref{eq:n1Integrated} (solid line) and the result from the Fermi Golden Rule approach (dashed line) agree very well, and the deviation between them is much smaller than 1 for all $|G_{1,1}|$. The parameters are $n_\text{th,1} = 100$, $n_\text{th,2} = 50$, $\omega_1/\gamma_1 = \omega_2/\gamma_2 = 10^5$, $\omega_1/\kappa = 0.01$, $\omega_2/\kappa = 0.02$, $\Delta_{0,1}/\kappa = -2$, and $C_{2,0} = 20000$. We note that ground state cooling is predicted also from the exact result, even for $|G_{1,1}| \sim |G_{2,0}|$.}
\label{fig:ExactSec4}
\end{figure}

While the frequency integral in \eqref{eq:PhononNumberExact} can be performed in the general case, it gives unwieldy expressions which do not offer much insight for the scenario studied here. We can nevertheless calculate the integral numerically, and the result for a particular set of parameters is shown in Fig.~\ref{fig:ExactSec4}. We observe that the deviation between the exact result and the Fermi Golden Rule result for the average phonon number is much smaller than 1 even for $|G_{1,1}| \sim |G_{2,0}|$.

\subsection{Physical realizations}
\label{sec:PhysRelSec4}

One example of a system for which the model in Eqs.~\eqref{hh},\eqref{hh2},\eqref{hh3} is relevant is again the membrane-in-the-middle setup. The optical cavity has many different modes, with different symmetry properties in the transverse dimensions of the cavity. Similarly, the membrane has different flexural eigenmodes, whose dependence on a transverse coordinate can be either symmetric or antisymmetric. 

The coupling rates $g_{i,jk}$ in the interaction Hamiltonian $H_{\text{int},i} = \sum_{j,k} \hbar g_{i,jk} (\hat{b}_i + \hat{b}^\dagger_i)\hat{a}^\dagger_j \hat{a}_k$ will be proportional to an integral over the transverse mode functions of the three modes involved. From this, it follows that $g_{i,jk} \neq 0$ requires that the coupling terms contain either zero or two modes that are anti-symmetric in a transverse coordinate. Thus, if e.g.~cavity mode 0 and mechanical mode 1 are symmetric in a transverse coordinate, and cavity mode 1 and mechanical mode 2 are antisymmetric in the same coordinate, the coupling terms in Eqs.~\eqref{hh2},\eqref{hh3} follow directly from symmetry considerations.

The model we used is also relevant to so-called optomechanical crystals, where a defect in a suspended photonic crystals can give rise to co-localized photon and phonon modes \cite{Eichenfield2009Nature_2,Safavi-Naeini2011NJP}, which can be viewed as a single optomechanical system. If two such defect sites are located close to each other, photon and phonon tunneling between the two sites can occur, in which case the system should be described in terms of delocalized supermodes \cite{Safavi-Naeini2011NJP}. For two identical defect sites, these supermodes will be either symmetric or antisymmetric in the coordinate along the axis connecting the two sites. Describing the on-site optomechanical interactions in terms of these supermodes will then give rise to the model we have studied in this section.

\section{Conclusion}
\label{sec:Conclusion}
We have presented a cooling scheme for optomechanical systems that allows for ground state cooling of mechanical motion when the resonance frequency of the mechanical oscillator is small compared to the linewidth of the optical cavity. This becomes possible with the inclusion of an auxiliary mechanical mode, which can be used to squeeze the photon number fluctuations in the cavity and thereby suppress the Stokes scattering that would otherwise hinder ground state cooling. The suppression can also be viewed as a result of OMIT, and the cooling scheme we presented has similarities with EIT cooling of the motion of atoms. In the case of a single cavity coupled to two mechanical oscillators, we showed that ground state cooling is possible if the auxiliary mode is in the resolved sideband regime. We also studied a modified setup with two optical cavity modes, and found that ground state cooling was then possible even if both mechanical modes were in the unresolved sideband regime. Finally, we commented on possible realizations of these schemes, for example with the membrane-in-the-middle setup or with optomechanical crystals.

\begin{acknowledgments}
The authors acknowledge financial support from The Danish Council for Independent Research under the Sapere Aude program (KB) and from the Academy of Finland (TO), as well as useful comments from Mika Sillanp{\"a}{\"a}, Andreas Nunnenkamp and Jack Harris. 
\end{acknowledgments} 


\end{document}